\documentclass[fleqn,usenatbib,onecolumn]{mnras}
\usepackage{graphicx}
\usepackage{epsfig}
\usepackage{amsmath}
\usepackage{multicol}
\usepackage{longtable}

\title[An Exploration of Nearby X-ray SNRs]{An Exploration of X-ray Supernova Remnants in the Milky Way and Nearby Galaxies}

\author[C. Albert and V. Dwarkadas]{Chris Albert$^{1}$ and Vikram V. Dwarkadas$^{1}$
\\
$^{1}$University of Chicago, Department of Astronomy and Astrophysics, 5640 South Ellis Avenue, Chicago, IL 60637, USA
}
\begin{document}
\maketitle
\begin{abstract}
    We probe the environmental properties of X-ray supernova remnants (SNRs) at various points along their evolutionary journey, especially the S-T phase, and their conformance with theoretically derived models of SNR evolution. The remnant size is used as a proxy for the age of the remnant. Our data set includes 34 Milky Way, 59 Large Magellanic Cloud (LMC), and 5 Small Magellanic Cloud (SMC) SNRs. We select remnants that have been definitively typed as either core-collapse (CC) or Type Ia supernovae, with well-defined size estimates, and a thermal X-ray flux measured over the entire remnant. A catalog of SNR size and X-ray luminosity is presented and plotted, with ambient density and age estimates from the literature. Model remnants with a given density, in the Sedov-Taylor (S-T) phase, are overplotted on the diameter-vs-luminosity plot, allowing the evolutionary state and physical properties of SNRs to be compared to each other, and to theoretical models. We find that small, young remnants are predominantly Type Ia remnants or high luminosity CCs, suggesting that many CC SNRs are not detected until after they have emerged from the progenitor's wind-blown bubble. An examination of the distribution of SNR diameters in the Milky Way and LMC reveals that LMC SNRs must be evolving in an ambient medium which is 30\% as dense as that in the Milky Way. This is consistent with ambient density estimates for the Galaxy and LMC. 
\end{abstract}


\begin{multicols}{2}
\section{INTRODUCTION}

 Supernovae (SNe) play an important role in galactic evolution. They transfer kinetic energy on the order of $10^{51}$ erg into the surrounding medium and expel the products of stellar nucleosynthesis \citep{Truran67}, thus enriching the medium with heavy elements which give rise to the next generation of stars \citep{Matteucci&Francois89}. Supernova remnants (SNRs), formed by the SN ejecta interacting with and sweeping up the ambient medium, can be observed for thousands of years. They form a third, hot phase of the interstellar medium \citep{mo77}.
 
 SNe are primarily divided into two categories, core-collapse (CC) and thermonuclear (Type Ia) SNe. A core-collapse SN occurs when the iron core of a massive star can no longer produce enough energy through fusion to counterbalance the star's gravitational energy \citep{Bruenn16, Burrows18}, resulting in the collapse of the star's outer layers onto the dense core, followed by the formation of a shock wave expanding outwards (see \citet{Janka12} for a review).
 Since massive stars continually lose mass throughout their evolution in the form of winds, the resulting SN shock will expand into the circumstellar wind-blown cavity created by mass-loss from the massive progenitor \citep{rac82b, Chev&Fran94}. 
 
 Thermonuclear SNe are thought to arise from white dwarf progenitors that may exceed the Chandrasekhar limit ($\approx 1.4 M_\odot $), thus becoming unstable \citep{Colgate&White66,Kerkwijk10}. The resulting subsonic deflagration wave eventually transitions to a supersonically expanding detonation wave that causes the white dwarf to explode \citep{Nomoto82,Vink12}. Since a white dwarf in isolation cannot exceed the Chandrashekhar mass to become unstable, the process is thought to be initiated by accreting material from a neighboring companion in a binary system, or by the merger of two white dwarfs \citep{Nomoto82,Webbink84,WoosleyKasen11}. Compared to CC remnants, white dwarfs do not have strong winds and do not lose much mass. Therefore the ambient medium that a Type Ia remnant evolves in can be assumed to be relatively constant. It is possible that the medium may be affected by mass loss from a companion star, although more observations are needed to evaluate the frequency of companion star types \citep{Bianco11}.

 SNR expansion is generally thought to proceed in 4 phases, although the details of the process are complicated, and some phases may be skipped under certain conditions \citep{Woltjer72,rac77,Jones98, dwarkadas11, Vink12}. Each phase can be characterized by a distinct relationship between the SNR radius and time of expansion. The remnant starts in the ejecta-dominated (ED) phase, when the mass ejected in the SN explosion dominates the dynamics of the SN shock wave. A reverse shock is formed that traverses back into the ejecta in a Lagrangian sense. As mass is continually swept up by the shock, the reverse shock eventually begins propagating inwards toward the center of the remnant. The time when the reverse shock reaches the center depends on the density profile of the ejected material and can vary considerably for different density profiles \citep{Dwarkadas&Chev98}. The Sedov-Taylor (S-T) phase is generally thought to begin after the remnant sweeps up an amount of material much larger than the ejected mass, and the shock radius increases with time as $R_{sh} \propto t^{2/5}$ (for a remnant expanding in a constant density medium) or $R_{sh} \propto t^{2/3}$ (for a remnant expanding in a wind with constant parameters). The shock expansion is considered to be adiabatic\footnote{Note that in reality a shock can never be adiabatic, although this terminology is commonly used.} \citep{Sedov59,Taylor50}.   Once the SN shock slows down and the energy radiated by the remnant becomes significant compared to the explosion energy, the SNR enters the radiative phase. If one assumes that cooling occurs just behind the shock wave, so that the mass inside the shock wave is basically all contained in this thin shell, then a momentum conserving solution can be derived, such that the shock radius $R_\text{sh} \propto t^{1/4}$. However, a more relevant solution  is the pressure-driven radiative shell \citep{om88}, where the interior mass is negligible but not the interior pressure. In that case radius $R_\text{sh} \propto t^{2/7}$ in the radiative stage. This should eventually transition to the momentum conserving solution, given enough time \citep{cmb88}. Finally, the shock wave essentially merges with the ambient medium and the SNR ceases to be distinguishable from its surroundings.

The evolution of a SNR over its lifetime has been theoretically described in the literature as outlined above. An important question that has not been adequately explored is whether the properties of observed SNRs match up to the theoretical expectations. SNRs may live for tens of thousands of years, while modern observations of SNRs have been carried out in the last 4-5 decades at best. In order to understand how their properties change over their lifetime, one needs to study a collective ensemble of objects, investigating observable characteristics that reflect the evolutionary phase. Such characteristics include the X-ray emission from SNRs, as well as their radius, as shown later in this manuscript (\S \ref{sec:theory} and \S \ref{sec:Lx-t}). Since the age of SNRs is often a debatable quantity, but their size can be measured (modulo the distance to the remnant) we use their size as a proxy for the age of the remnant. While there are several caveats inherent in their calculation, which are mentioned in the appropriate sections, these parameters can be described by relatively simple equations.  Thus observing the variations in the X-ray emission with radius from an ensemble of SNRs can  shed light on the evolutionary properties of SNRs.

{\em In this paper our goal is to evaluate whether the properties of X-ray SNRs are consistent with the theoretical models of SNR evolution, and reflect accurately the evolutionary phase they exist in.}  Using phase appropriate equations for the SN expansion, coupled with the thermal X-ray luminosity and its variation with time, we attempt to accurately characterize Galactic and nearby supernova remnants, evaluate how their thermal X-ray emission evolves with time, and whether it corresponds to the evolutionary sequence that is generally proposed in textbooks and  was outlined earlier. 

Our study requires knowledge of the X-ray flux of SNRs. We primarily focus on remnants that are dominated by thermal emission, but for the sake of completeness also include a few well-studied remnants that show strong nonthermal components, such as Tycho's SNR and SN 1006. Some SNRs may contain pulsar-wind nebulae in the center, which can produce non-thermal X-ray emission \citep{Berezhko04}. We outline general trends in the evolution of Galactic SNRs, and compare these to remnants in the Large and Small Magellanic Clouds (LMC and SMC), the former of which were studied earlier by \citet{Ou18}. The well calibrated distance to the LMC considerably reduces the uncertainty in the remnant properties, especially the X-ray luminosity $L_x$. The much larger uncertainty in Galactic distances is reflected in the properties of the Galactic remnants, thus resulting in large error bars in the properties. In some cases, our investigations can help to reduce the error bars. 

In \S \ref{sec:data} we list the SNRs in this study and their properties. We consider 34 Galactic remnants, 59 LMC remnants, and 5 SMC remnants, all of which have been characterized as core-collapse or Type Ia. \S \ref{sec:theory} provides an overview of SNR evolution, while \S \ref{sec:Lx-t} discusses the evolution of the X-ray luminosity of SNRs with time. In \S \ref{sec:results} we list our findings for Galactic core-collapse and Ia remnants, comparing them with those in the LMC and SMC. We plot the SNR size against the X-ray luminosity and offer explanations for the visible trends. We also demonstrate how the results offer a method to compare differing distance estimates for galactic remnants. The implications of different distance estimates are evaluated in \S \ref{subsec:compare}. We compare the average densities of the Milky Way and LMC in \S \ref{subsec:total} by studying all SNRs with a size estimate within them. \S \ref{sec:concl} summarizes the results and conclusions.

\section{SNR DATA} \label{sec:data}

The Galactic SNR population suffers from uncertainties in distance measurements, as well as dust in the galactic plane obscuring more distant remnants. The relative proximity of Galactic SNRs, compared to those in other galaxies, makes the distance uncertainty a significant concern. This is amplified by the fact that the luminosity depends on the square of the distance $d$, while the conversion from angular to linear size depends on $d$. 

X-ray luminosities used in this paper are primarily taken from the Chandra Supernova Remnant Catalog\footnote{https://hea-www.harvard.edu/ChandraSNR/} and the papers listed within it. We only use those remnants for which the flux across the entire SNR, as well as the SNR type, was listed. Since the catalog has not been updated for many years, more recently published distance estimates are used where available. The SNR Catalog\footnote{http://snrcat.physics.umanitoba.ca/} provided by \citet{FerrandSafi-Harb12} was used to identify papers providing updated and additional properties. The source of the values found in Table \ref{table:Gal} is denoted with a superscript. Other papers that discuss measurements are sometimes also included. The Chandra Catalog is indicated with the superscript $^*$, and the distances for the diameter estimates are from the same sources as those used for $L_x$. For SNRs where a revised distance estimate exists compared to the paper measuring X-ray flux, we recalculate the diameter and luminosity using the new estimate. We use the same approximation for diameter ($D$) as used by \citet{Ou18} for LMC remnants, that of averaging the major and minor axes of each remnant's centroid.

The type, age, and ambient density of the studied remnants are given in Table \ref{table:Gal}.  Different methods of determining distance can yield different age and density values, so the ranges given in the table are based on a combination of the error limits given in the literature and the differing estimates. If part of a remnant is interacting with a dense molecular cloud, this region of high ambient density may not be reflected in the tabulated $n_0$, unless it was already included in the published $n_0$. The highly asymmetric shape of G350.1-0.3 suggests we may only be receiving X-ray emission from a cloud interaction, thus its tabulated ambient density reflects the density of the cloud \citep{Gaensler08, Borkowski20}. Cloud interactions are discussed in Section \ref{sec:results} if they appear to have had a significant effect on specific remnants. 

The 5 SMC remnants included in this study are presented in Table \ref{table:SMC}. Their properties are extracted from the Chandra Supernova Remnant Catalog and papers listed therein, with the diameters found by averaging the axes of their centroids, and with age and ambient density estimates from the literature. The LMC SNR luminosity and diameter estimates we use are given in Table 1 of \citet{Ou18}. This combines X-ray luminosity data from \citet{Maggi16} with size estimates cross-checked between \citet{Desai10,Badenes2010,Bozzetto17}.

\end{multicols}

\begin{longtable}{cccccc p{3cm}}
\caption[]{Galactic X-Ray SNRs}\\
\hline
        Name & SN Type & Diameter (pc) & $L_x$ (erg s$^{-1}$) & Age (yr) & $n_0$ (cm$^{-3}$) & References\\ \hline
        G0.9+0.1 & CC & 3.05 & 2.55e+35$^{*,4}$ & 1100-6800$^5$ & ~ & \citet{Gaensler01}$^4$; \citet{Mereghetti98}$^5$; \citet{Sidoli04} \\ \hline
        G1.9+0.3 & 1A & 4.33 & 6.74e+34$^{*,6}$ & 120$^{7}$ & 0.02-0.2$^{7,8}$ & \citet{Reynolds08}$^6$; \citet{Carlton11}$^7$; \citet{Borkowski14}$^8$ \\ \hline
        Kepler SNR & 1A & 5.96 & 2.04e+36$^{*,9}$ & 420$^{10}$ & 5.8-9.2$^{10}$ & \citet{Reynoso99}$^9$; \citet{SunChen19}$^{10}$ \\ \hline
        G11.2-0.3 & CC & 6.47 & 1.19e+37$^{*,11}$ & 1600$^{11,12}$ & ~ & \citet{Green88}$^{11}$; \citet{Koo07}$^{12}$ \\ \hline
        G15.9+0.2 & CC & 11.87 & 8.24e+36$^{*,14}$ & 2000-6000$^{13,14}$ & 0.561-0.946$^{13}$ & \citet{Sasaki18}$^{13}$; \citet{Reynolds06}$^{14}$ \\ \hline
        Kes 73 & CC & 8.97 & 9.53e+36$^{*,15}$ & 2000-4000$^{16}$ & 0.8-2$^{15,16}$ & \citet{GotthelfVasisht97}$^{15}$; \citet{BorkowskiReynolds17}$^{16}$ \\ \hline
        3C 391 & CC & 15.59 & 4.75e+36$^{*,17}$ & 4000-17000$^{17,18,19}$ & 0.07-0.4$^{17}$ & \citet{Chen04}$^{17}$; \citet{Sato14}$^{18}$; \citet{SuChen05}$^{19}$ \\ \hline
        Kes 79 & CC & 18.33 & 1.02e+36$^{*,22}$ & 4400-6700$^{20}$ & 0.13-8.0$^{20,21}$ & \citet{Zhou16}$^{20}$; \citet{Giacani09}$^{21}$; \citet{Sun04}$^{22}$ \\ \hline
        G38.7-1.4 & CC & 9.31 & 7.29e+33$^{*,23}$ & 13900-15100$^{23}$ & 0.05-0.1$^{23}$ & \citet{Huang14}$^{23}$ \\ \hline
        3C 396 & CC & 15.13 & 1.43e+36$^{*,25}$ & 6000-7100$^{24,25}$ & 1.0$^{24}$ & \citet{Lee09}$^{24}$; \citet{Olbert03}$^{25}$ \\ \hline
        3C 397 & 1A & 18.85 & 1.55e+37$^{*,27}$ & 1350-5300$^{26,27}$ & 2.0-5.0$^{26}$ & \citet{LeahyRanasinghe16}$^{26}$; \citet{Safi-Harb05}$^{27}$ \\ \hline
        W49B & 1A & 11.52 & 4.57e+37$^{*,30}$ & 1000-6000$^{28,30,31}$ & 2.0-24$^{30,31}$ & \citet{Zhu14}$^{28}$; \citet{Miceli08}; \citet{ZhouVink18}$^{30}$; \citet{Yamaguchi12}$^{31}$ \\ \hline
        G67.7+1.8 & CC & 29.09 & 5.23e+34$^{*,32}$ & 5000-13000$^{32}$ & 0.015-0.25$^{32}$ & \citet{HuiBecker09}$^{32}$ \\ \hline
        Cas A & CC & 5.64 & 2.84e+37$^{*,33}$ & 340$^{33}$ & 2-6$^{34}$ & \citet{Reed95}$^{33}$; \citet{Kim08}$^{34}$; \citet{LamingHwang03} \\ \hline
        Tycho & 1A & 10 & 1.37e+36$^{*,37}$ & 450$^{35}$ & 0.2-1$^{35,36,37}$ & \citet{wtdp20}$^{35}$; \citet{MorlinoCaprioli12}$^{36}$; \citet{Giordano12}$^{37}$ \\ \hline
        Puppis A & CC & 25.45 & 1.2e+37$^{40}$ & 3700-4450$^{39,40}$ & 0.15-10$^{38}$ & \citet{Hwang05}$^{38}$; \citet{Becker12}$^{39}$; \citet{Dubner13}$^{40}$ \\ \hline
        G272.2-3.2 & 1A & 20.36 & 3.97e+35$^{*,41}$ & 6000-11000$^{41}$ & 0.46-1$^{41}$ & \citet{Greiner94}; \citet{McEntaffer13}$^{41}$ \\ \hline
        MSH 11-61A & CC & 28.51 & 4.71e+35$^{*,42}$ & 10000-30000$^{42,43}$ & 0.7-1$^{43}$ & \citet{Slane02}$^{42}$; \citet{Garcia12}$^{43}$ \\ \hline
        MSH 11-54 & CC & 12.57 & 5.73e+36$^{*,44}$ & 3000$^{45}$ & .875-2.75$^{46}$ & \citet{Park02}$^{44}$; \citet{Winkler09}$^{45}$; \citet{GhavamianWilliams16}$^{46}$ \\ \hline
        G306.3-0.9 & 1A & 5.53 & 7.47e+36$^{*,47}$ & 2500-6000$^{47,48}$ & .53-1.08$^{47}$ & \citet{Sawada19}$^{47}$; \citet{Reynolds13}$^{48}$ \\ \hline
        G308.3-1.4 & CC & 17.10 & 1.38e+35$^{*,49}$ & 5000-7500$^{49}$ & .165-.22$^{49}$ & \citet{PrinzBecker12}$^{49}$; \citet{DeHorta13} \\ \hline
        G327.1-01.1 & CC & 28.80 & 1.44e+35$^{*,50}$ & 17400$^{50}$ & 0.12$^{50}$ & \citet{Temim15}$^{50}$ \\ \hline
        Kes 27 & CC & 15.13 & 3.64e+35$^{*,51}$ & 3500-8400$^{51,52}$ & 0.4$^{51}$ & \citet{Chen08}$^{51}$; \citet{Seward96}$^{52}$ \\ \hline
        SN1006 & 1A & 16.6 & 3.4e+34$^{53}$ & 1015$^{52,53}$ & 0.02-0.3$^{52,53}$ & \citet{Uchida13}$^{53}$; \citet{Dubner02}$^{54}$ \\ \hline
        G330.2+1.0 & CC & 16.00 & 8.41e+34$^{*,55}$ & 800-1200$^{55,56}$ & 0.1-.25$^{55,57}$ & \citet{Park09}$^{55}$; \citet{Borkowski18}$^{56}; $\citet{Williams18}$^{57}$ \\ \hline
        Kes 32 & CC & 26.18 & 4.55e+35$^{*,58}$ & 3000$^{58}$ & 0.6-1.3$^{58}$ & \citet{Vink04}$^{58}$ \\ \hline
        G337.2-0.7 & 1A & 9.89 & 7.11e+36$^{*,59}$ & 750-7000$^{59}$ & 0.63-1.6$^{59}$ & \citet{Rakowski06}$^{59}$; \citet{Yamaguchi_Badenes14} \\ \hline
        Kes 41 & CC & 26.18 & 1.3e+35$^{60}$ & 1500-16000$^{61}$ & 0.2-0.4$^{61}$ & \citet{Combi08}$^{60}$; \citet{Zhang15}$^{61}$ \\ \hline
        G340.6+0.3 & 1A & 24.92 & 3.31e+36$^{*,62}$ & 2600$^{62}$ & ~ & \citet{Caswell83}$^{62}$; \citet{Dubner96}; \\ \hline
        G344.7-0.1 & 1A & 24.43 & 2.22e+37$^{*,64}$ & 3000-6000$^{63,64}$ & 0.2-0.3$^{63,64}$ & \citet{Combi10}$^{63}$; \citet{Giacani11}$^{64}$ \\ \hline
        CTB 37A & CC & 27.20 & 2.53e+35$^{*,65}$ & 10000-24000$^{65,66}$ & ~ & \citet{Yamauchi_Minami14}$^{65}$; \citet{Maxted13}$^{66}$ \\ \hline
        G349.7+0.2 & CC & 15.36 & 1.89e+37$^{*,68}$ & 2800$^{68}$ & 5-25$^{68}$ & \citet{Frail96}$^{67}$; \citet{Lazendic05}$^{68}$; \citet{Yasumi14} \\ \hline
        G350.1-0.3 & CC & 4.97 & 3.97e+36$^{*,69}$ & 600-1200$^{69,70,72}$ & 9.375-600$^{69,71}$ & \citet{Gaensler08}$^{69}$; \citet{Lovchinsky11}$^{70}$; \citet{Yasumi14}$^{71}$; \citet{Borkowski20}$^{72}$ \\ \hline
        G352.7-0.1 & CC & 11.64 & 1.99e+37$^{*,73}$ & 1600-4700$^{73,74,75}$ & 0.07-1.225$^{73,75}$ & \citet{Giacani09}$^{73}$; \citet{Kinugasa98}$^{74}$; \citet{Pannuti14}$^{75}$ \\ \hline

\label{table:Gal}
\end{longtable}

\begin{longtable}{cccccc p{3cm}}
\caption[]{SMC X-Ray SNRs}\\
\hline
        Name & SN Type & Diameter (pc) & $L_x$ (erg s$^{-1}$) & Age (yr) & $n_0$ (cm$^{-3}$) & References\\ \hline
        IKT 2 & CC & 22.7 & 1.74e+35$^{*,76}$ & 5400$^{76}$ & 0.11$^{76}$ & \citet{Heyden04}$^{76}$\\
        \hline
        IKT 6 & CC & 43.6 & 2.98e+36$^{*}$ & 14000-17000$^{76,77,78}$ & 0.04-0.5$^{76,78}$ & \citet{Hendrick05}$^{77}$; \citet{Schenck14}$^{78}$\\
        \hline
        IKT 18 & CC & 33.2 & 2.14e+35$^{*}$ & 6000$^{79}$ & 0.03-0.05$^{76,79}$ & \citet{Yokogawa02}$^{79}$\\
        \hline
        IKT 22 & CC & 12.8 & 2.53e+37$^{*,80}$ & 1450-2700$^{76,81}$ & 0.42$^{76}$ & \citet{Gaetz00}$^{80}$ \citet{Finkelstein06}$^{81}$\\
        \hline
        IKT 23 & CC & 50.6 & 1.39e+37$^{*}$ & 18000-19000$^{76,82}$ & 0.04-0.2$^{76,82}$ & \citet{Park03}$^{82}$\\
        \hline
        
\label{table:SMC}
\end{longtable}

\begin{multicols}{2}

\section{SNR EVOLUTION} \label{sec:theory}
The expansion of a SN into the surrounding medium leads to a forward shock expanding into the ambient medium, and a reverse shock expanding back into the ejecta, separated by a contact discontinuity. For a SNR with an outer power-law ejecta density profile $\rho_E = \left(r/tg\right)^{-n} t^{-3}$, with constant $g$, evolving in a stationary ambient medium with a density profile $\rho_A = qr^{-s}$, with constant $q$, the evolution of the contact discontinuity can be described by a self-similar solution \citep{rac82a}
\begin{equation}
    \label{eqn:chev_rad}
    R_{\text{CD}} = \left( \frac{A g^n}{q}\right)^{1/(n-s)} t^{(n-3)/(n-s)},
\end{equation}
with the constant $A$ tabulated for given values of $s$ and $n$ in \citet{rac82a,Chev&Fran94}. The self-similar solution exists provided that $s<3$, $n\geq 5$, and the reverse shock is still expanding within the power-law portion of the ejecta.

A remnant may spend a large part of its visible life in the Sedov-Taylor (S-T) phase. While the S-T phase may be reached while the remnant is expanding in a wind, it is uncommon, since the swept-up mass must exceed the SN ejecta mass by a factor of 20-30 before the S-T phase can begin \citep{Dwarkadas&Chev98}, and the mass in the wind is usually not large enough. Therefore it is convenient to assume that the remnant is interacting with the interstellar medium (ISM) when it is in the S-T phase.

The relationship between the radius ($R_\text{ST}$) and time of evolution $t$ in the S-T phase was derived independently by both Sedov and Taylor, and therefore known as the Sedov-Taylor relation. It can be written \citep{Taylor50} as:
\begin{equation}
    \label{eqn:st_rad}
    R_{\text{ST}} = \left( \frac{\xi(\gamma) E}{\rho_0} \right) ^{1/5} t^{2/5},
\end{equation}
where $E$ is the explosion energy, $\rho_0$ the mass density of the medium into which it propagates, and $\xi(\gamma)$, a function of the ratio of the specific heat of the gas. The latter has a value of $\xi(\gamma)$=2.206 for a non-relativistic, monotonic gas \citep{Vink12}.

The shock temperature in our model S-T remnants varies with the radius as 
\begin{equation}
    \label{eqn:st_shock_temp}
    T_{\text{shock}} = \left( \frac{E}{n_0} \right)  R_{\text{ST}}^{-3} \cdot 10^{10} \text{K},
\end{equation}
where $E$ is in units of 10$^{51}$ ergs, $n_0$ in cm$^{-3}$, and $R_{\text{ST}}$ in pc \citep{SewardCharles10}. This is the post-shock temperature, calculated assuming temperature equilibration between electrons and ions, which may not necessarily be true. Since supernova shocks are collisionless, energy in the shock is primarily transferred to the ions, and the electron temperature is correspondingly lower than the ion temperature. As the shock velocity decreases, collisions between electrons and ions increase, and the electron temperature begins to approach the ion temperature, as shown by \citet{glr07} in their analysis of the forward shock in Balmer-dominated SNRs. While the assumption of a higher electron temperature could affect the calculation of the X-ray luminosity, the effect is not large in the Sedov phase which we focus on.  In Tycho's SNR, there is clear evidence of efficient collisionless heating of electrons, albeit at the reverse shock \citep{yamaguchietal14} where the physical conditions may be different compared to the forward shock. Nonetheless, this shows that diverse plasma processes at shocks can possibly bring the electron temperature closer to the ion temperature than would be possible by Coulomb collisions alone.

\section{X-ray luminosity of SNRs with time} 
\label{sec:Lx-t}
The thermal X-ray emission from SNe is comprised of thermal bremmsstrahlung combined with line emission from various elements. The luminosity $L_x$ can then be written as
\begin{equation}
    \label{eqn:lumx}
    L_x = V \Lambda n^2,
\end{equation}
where $V$ is the volume of the emitting region, $\Lambda \propto T^{-1/2}$ for temperatures $10^5 K \leq T \leq 10^{7.5} K$ and $\Lambda \propto T^{1/2}$ for $T\geq 10^{7.5} K$ is an approximation used by \citet{Truelove&Mckee99} for the cooling function shown in \citet{Raymond76}, and $n$ is an average number density within the remnant. For the S-T phase SNRs we take $n = n_0$ under the assumption that the ejecta mass is insignificant compared to the mass swept up by the shock, so the total mass within the SNR's internal volume is equal to the mass swept up in the ISM. Although we ignore the mass of the ejecta compared to the total mass, it is possible that the X-ray luminosity of the ejected material may be high in proportion to its mass, especially in younger SNRs, since the ejecta are mainly composed of heavy elements. We return to this point in later sections in relation to Tycho's SNR and the SNR Cas A.

A CC remnant will evolve in the wind medium created by the massive star progenitor. Close to the SN, this will essentially take the form of a freely expanding wind. A wind with a constant mass-loss rate ${\dot{M}}_w$ and wind velocity $v_w$ (steady wind) has a density that decreases as 
$\rho = {\dot{M}_w}/({4 \pi v_w r^2})$. Assuming a cooling function that goes as temperature $T^{1/2}$, using equation (\ref{eqn:lumx}), and the Rankine-Hugoniot conditions which state that $T \propto v_{sh}^2$, it can be shown that the X-ray emission decreases with time as $L_x \propto t^{-1}$ \citep{flc96, dg12}. Since the remnant first expands in the stellar wind, the X-ray luminosity is expected to continuously decrease. The wind ends in a wind termination shock, followed by a more or less constant low density region, and then a dense shell \citep[i.e. a wind bubble;][]{weaveretal77, Dwarkadas05}. The X-ray emission should begin to increase in the constant density region according to equation (\ref{eqn:lumx}), followed by a significant increase as the shock impacts the dense shell.

Type Ia remnants are generally not considered to be evolving in a wind but in a constant density medium. The X-ray luminosity predicted by equation (\ref{eqn:lumx}), for a constant density $n_0$, will be increasing in the ED stage, since the volume is increasing as $r^3$ and the cooling function $\Lambda \propto T^{1/2} \propto  v_{sh} \propto r/t$. The time dependence is a function of the ejecta profile and is difficult to deduce, but the X-ray luminosity will be of order $L_x \propto t^2$. Therefore it is not unusual to see small size remnants of Type 1a in the LMC since their X-ray luminosity is increasing with time.

Once the remnant is expanding in the ISM, the temperature decreases to below a few keV, while the density is constant. Equation (\ref{eqn:lumx}) can then be used to show that the X-ray luminosity will increase with time \citep{rd20} as $L_x \propto t^{1.8}$. Thus, in core-collapse remnants, the luminosity will initially decrease with time, followed by perhaps an increase (but for a low value and in a low density bubble) and then a large increase if the SN shock impacts a dense shell \citep{Dwarkadas05}. When the SN shock wave emerges from the bubble and is interacting with the ISM in the S-T phase, its luminosity will first decrease since it is expanding in a lower-density ISM, and then steadily increase with time. 

In the case of Type Ia remnants, the luminosity evolution in the S-T phase is somewhat similar to that in the ED phase, since the density distribution may not change. However the radius, and therefore volume, increases at a slower rate with time, as given by equation (\ref{eqn:st_rad}), while the cooling function goes with temperature approximately as $T^{-1/2}$. Type Ia remnants should not show a large difference between the luminosity evolution in the ED and S-T stage. 

We note that the X-ray luminosities of the remnants listed here are measured mainly with Chandra and XMM-Newton, and are usually in the 0.3-8 keV band, or somewhere in the range between 0.3-10 keV. However, the total X-ray luminosity, which is what is referred to in equation (\ref{eqn:lumx}), could be much higher if the X-ray temperature is outside this range. This is mainly a problem for young remnants with strong fast shocks, where the post-shock temperature can lie outside the Chandra/XMM-Newton range. By the time that remnants are in the S-T phase, which are the majority of the studied remnants, the temperature is generally a few keV, and we would expect that most of the X-ray emission falls in the Chandra or XMM-Newton band, so the majority of our calculations are justified. Still, we must keep this caveat in mind, especially for the younger remnants, of which there are a few.

\section{RESULTS} \label{sec:results}

\begin{figure*}
\includegraphics[width=\textwidth,angle=0]{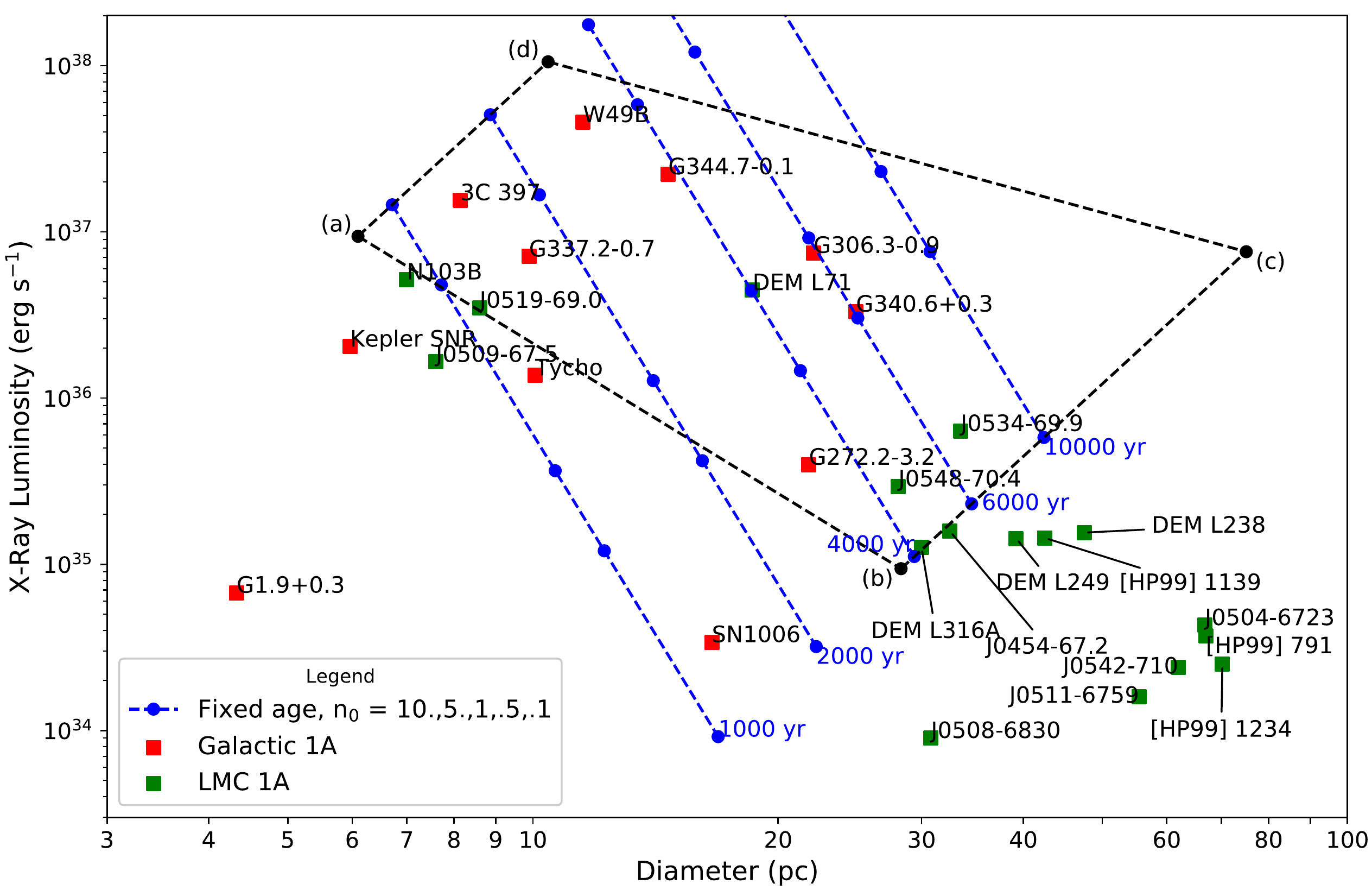}
\caption{The X-Ray luminosity and diameter for Galactic and LMC Type Ia remnants. The black quadrilateral demarcates the region where Type Ias lie in the S-T phase, with ISM density 0.1 cm$^{-3}$ $< n_0 <$ 10 cm$^{-3}$ and initial energy $E=10^{51}$ erg. The blue lines show the position of a remnant in the S-T stage, taking the average internal density to be equal to the ambient density and the temperature of X-ray emitting regions to be the post-shock temperature.
\label{fig:1A}}
\end{figure*}

\begin{figure*}
\includegraphics[width=\textwidth,angle=0]{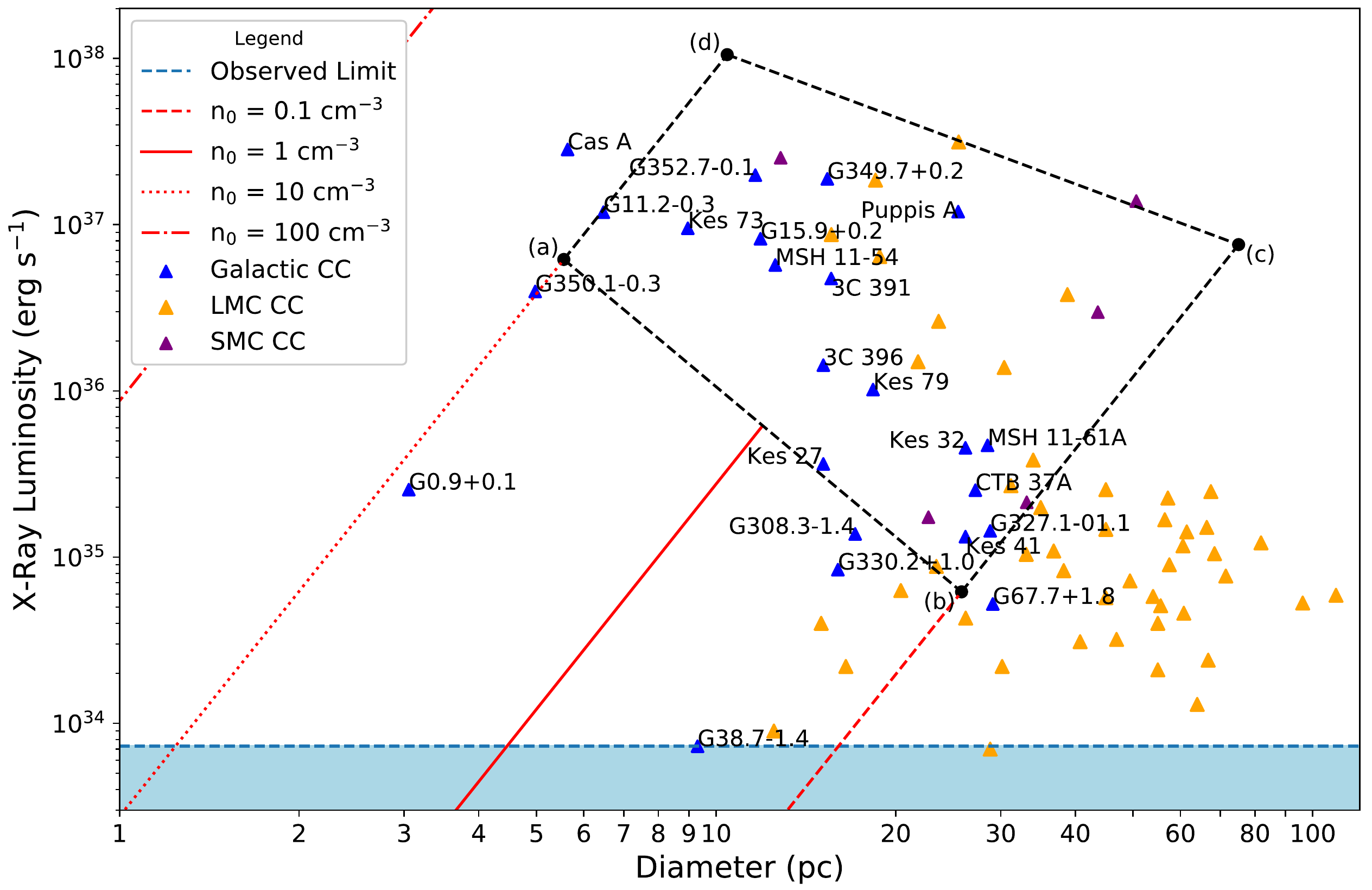}
\caption{The X-Ray luminosity and diameter for Galactic, LMC, and SMC CC remnants. The blue band shows the lower limit of observed luminosities. The red lines show the X-ray luminosity versus size for constant density, with an initial explosion energy of $10^{51}$ erg.
\label{fig:CC}}
\end{figure*}

Figures \ref{fig:1A} and \ref{fig:CC} show the X-ray luminosity of Ia and core-collapse remnants in the Galaxy and Magellanic clouds against their diameter. In both figures, the black dashed lines form the boundary of the region where remnants evolving in a constant density ISM would lie if they were in the Sedov-Taylor phase and were evolving in a medium with ambient density between 0.1 cm$^{-3}$ and 10 cm$^{-3}$. The line between points (a) and (b) denotes the line marking the entry of SNRs into the S-T phase, while the line between (c) and (d) demarcates the end of the S-T phase and the beginning of the radiative phase. (a) and (d) correspond to remnants evolving in a medium with $n_0=10$ cm$^{-3}$, and (b) and (c) to SNRs evolving in a medium with $n_0=0.1$ cm$^{-3}$. The lines connecting (a) to (d) and (b) to (c) represent the path a SNR with ambient density 10 cm$^{-3}$ and 0.1 cm$^{-3}$, respectively, would follow as it progresses through the S-T phase. An explosion energy of $E = 10^{51}$ erg, and (in Figure \ref{fig:1A}) a progenitor mass $M_e = M_\text{Ch} = 1.4 M_\odot$ (the Chandrasekhar mass), is assumed. 

For Type Ia remnants, the exponential SNR density profile \citep{Dwarkadas&Chev98} was used, which leads to remnants having a larger size when entering the S-T phase as opposed to those with a power law density profile. Normalization parameters for radius and age can be written as
\begin{equation}
    R' \approx 2.19\left( \frac{M_e}{M_\text{Ch}} \right) ^{1/3} n_0^{-1/3} \text{pc}
\end{equation}
\begin{equation}
    T' \approx 248 E_{51}^{-0.5} \left( \frac{M_e}{M_\text{Ch}} \right) ^{5/6} n_0^{-1/3} \text{yr},
\end{equation}
such that the evolution of the SNR can be described in terms of dimensionless values $r' = r/ R'$, $t' = t/T'$. \citet{Dwarkadas&Chev98} find that the reverse shock propagates back to the center of the remnant at approximately $r'\approx 3$ and $t' \approx 7.5$, which we use as the starting point for the S-T phase. 

Core-collapse SNRs are initially expected to evolve in the stellar wind of a massive star with an ambient density profile that varies as $s=2$. This introduces additional parameters such as the progenitor's mass loss rate $\dot{M}$ and wind velocity $v_w$ to describe the density $\rho \propto \dot{M}/(v_w)$. These parameters are not well calibrated and may vary up to several orders of magnitude throughout a star's life, especially when a star transitions from a main-sequence to a post-main-sequence phase. They may also differ between individual stars of similar mass. In order to draw a boundary that is generally acceptable, some simplifying assumptions are used here. As shown by several authors starting from \citet{gull73}, and quantified in \citet{Dwarkadas&Chev98}, the Sedov stage is only reached when the swept-up mass exceeds the ejected mass by a factor of 20-30. Thus even for 5 M$_{\odot}$ of ejected material, one would need to sweep up 100-150 solar masses before the remnant can be expected to be in the Sedov stage. Furthermore, as shown by \citet{sukhboldetal16}, about 90\% of SNe arise from massive stars below about 20 M$_{\odot}$. These stars do not lose a lot of mass, and we could estimate that the wind mass would be a small fraction of the swept-up mass of 100-150 M$_{\odot}$ needed. Therefore, it is reasonable to assume that the remnant would generally have to be expanding within the ISM for a substantial period of time before the S-T phase is reached. We can then calculate the boundary using the solution for a constant density medium, keeping the above approximations in mind. Simplifying the plotted S-T boundary using $s=0$ allows for comparison to model remnants with a known age and ambient density. The unified \citet{Truelove&Mckee99} solution for the dimensionless S-T transition time and characteristic scaling factor for a remnant in a constant density medium is
\begin{equation}
    t_{\text{ST}}' \approx 0.495 \left[ \frac{5}{3} \left(\frac{3-n}{5-n}\right)\right]^{1/2}
\end{equation}
\begin{equation}
    T' = E^{-1/2} M_{\text{ej}}^{5/6} \rho_0^{-1/3},
\end{equation}
where $\rho_0 = m_{\text{H}} n_0$. We use a value of ejecta mass $M_{\text{ej}} = 5 M_\odot$. Likewise, since $t_{\text{ST}}'$ decreases asymptotically for $n>5$, we use $n=12$ so that most reasonable SNR ejecta profiles \citep{rac82a} will be encompassed within our S-T region.

The upper-right boundary line in both figures represents the time at which remnants leave the S-T stage and enter the radiative stage according to the \citet{Truelove&Mckee99} solution of

\begin{equation}
    R_{\text{rad}} \approx 14.0 E_{51}^{2/7}n_0^{-3/7} \zeta_m^{-1/7} \text{pc} \label{eqn:Rrad}
\end{equation}
\begin{equation}
    t_{\text{rad}} \approx 13300 E_{51}^{3/14}n_0^{-4/7} \zeta_m^{-5/14} \text{yr}.
\end{equation}
The dimensionless constant $\zeta_m$ accounts for metallicity differences between the current SNR environment and that in which the cooling function $\Lambda$ was fit. For simplicity, we assume solar abundances and take $\zeta_m = 1$ \citep{Truelove&Mckee99,Raymond76}.

The blue dashed lines in Figure \ref{fig:1A} represent the radius of the remnant at a given age, for various densities (assuming a fixed explosion energy of $10^{51}$ erg), as given by equation (\ref{eqn:st_rad}) and taking $\rho_0 = m_\text{H} n_0$. Each solid dot along the lines represents the location of the SNR at a given density, going from $n_0 = 10$ cm$^{-3}$ at the lower end of the size range, to $n_0 = 0.1$ cm$^{-3}$ at the higher end.

\subsection{Type Ia Remnants} 
Luminosities of most Galactic Type Ia SNRs appear to be fairly compatible with S-T models in a constant density ISM, with some exceptions.  SN 1006 appears to exceed the age estimate of 1000 years by a few hundred years, and the plot indicates an ambient density of $\sim$ 0.2 cm$^{-3}$. In reality the density around the remnant is known to vary widely, but is generally $<$ 0.1 cm$^{-3}$ around most regions except in the NW region where it could be higher \citep{abd07, Uchida13}.  The position of SN 1006 indicates that if the luminosity were to arise from purely thermal emission, then it would need a high average density of $\sim 0.2$ cm$^{-3}$ in the medium around it, higher than what is actually deduced for the surrounding medium. This is consistent with the fact that the emission from the SNR is predominantly non-thermal synchrotron emission, and the density around it is lower than would be expected from a thermal model. 

Tycho's SNR is only about 450 years old, but its position on the plot indicates an age exceeding 1000 years and a density 2-3 cm$^{-3}$. A review of various estimates for the density around Tycho's SNR is given in \citet{wtdp20}. The ambient density is generally calculated to lie between 0.2-1 cm$^{-3}$. Thus both the age estimate as well as the density are high given the current estimated properties.  The inference from its density could be similar to that derived for SN 1006 -  if the entire luminosity arises from thermal emission, it would need a higher density than is actually observed around it. A possibility is that some fraction of the emission may be non-thermal. \citet{Warren05} estimate that up to 60\% of the 4-6 keV continuum emission must be non-thermal. Furthermore, a large contribution to the emission comes from the reverse-shocked ejecta, as has been noted by \citet{hg97} and \citet{micelietal15}. Tycho's SNR also has bright knots of ejecta that contribute to its X-ray luminosity \citep{yamaguchietal17, williamsetal20}. Thus, although the total ejecta mass is $\le 1.4 M_{\odot}$, the ejecta in Tycho's SNR make an outsize contribution to the X-ray luminosity compared to their mass, resulting in an overall higher luminosity. A possibility for the large size of Tycho at its age is that the density in which it is expanding was lower at some point in the past than it is today, leading to a size (and corresponding age) that is larger than what would be expected at the current density. This was one of the suggestions made by  \citet{Dwarkadas&Chev98} to explain Tycho's X-ray emission profile, and is also echoed by \citet{chiotellisetal13} and \citet{yamaguchietal14}.  Its position on the plot indicates that although Tycho is younger than SN 1006, it is in a more evolved state than SN 1006, presumably due to a higher ambient density.

DEM L71's age is estimated to be around 4000 years \citep{hughesetal03,ghavamianetal03,vanderheydenetal03,franketal19}, with a density that varies but could be around 1--2 cm$^{-3}$. These parameters are consistent with its position on the plot. 

It is not certain whether W49B is a Type Ia remnant or not. It was thought to be a core-collapse remnant for many years \citep{lopezetal13}, but was recently classified as a Type Ia remnant by \citet{ZhouVink18}, a designation supported by \citet{siegeletal20b}. Although its age range is large, and thus its position is consistent with its age, the density appears to be higher than expected. W49B is known to be interacting with a dense cloud on one side, which could lead to an increase in X-ray emission, and thus a higher average value of the density \citep{ZhouVink18,siegeletal20b}.

3C 397 is classified as a Type Ia SNR. There is some debate about this. Based on a comparison between the Fe K$\alpha$ line centroid energies, fluxes, and elemental abundances of intermediate-mass and heavy metals (Mg to Ni) to Type Ia and CC hydrodynamical model predictions, \citet{martinezetal20} conclude that it is a Type Ia remnant. In contrast, \citet{siegeletal21} have shown that the Fe K$\alpha$ line centroid energy varies with position around the remnant, and cannot be used to clearly delineate the type. The remnant lies near the lower end of the suggested age range of 1300-5000 years. The estimated density at its position is consistent with the maximum density from the range of 2-5 cm$^{-3}$ given by \citet{LeahyRanasinghe16}.

The position of Kepler's SNR matches its age of around 420 years and suggests a high density of 4-5 cm$^{-3}$. The high average density may be indicative of expansion in a higher density medium in the past, perhaps a stellar wind, compatible with detailed hydrodynamical modeling carried out by \citet{patnaudeteal12}.

\citet{dickel20} suggests that the SNR N103B in the LMC is actually composed of two SNRs, a smaller bright one and a newly identified large one. The X-ray luminosity mentioned here would then be due to both remnants, thus making it difficult to characterize its parameters. \citet{yamaguchietal21} on the other hand suggest that the SNR is expanding into an hourglass shaped cavity and thus forming bipolar bubbles of ejecta. In either case, there is clear evidence of density inhomogeneities around the remnant. Densities around the remnant can in some regions be as high as 1500 cm$^{-3}$ \citep{yamaguchietal21} so the high average density of N103B reflected in its position in the plot is hardly surprising.

We find that the Galactic population of X-ray SNRs has very few large ($> 30$ pc) Type Ia remnants when compared to the LMC population. Searching the literature for Galactic remnants observed at other wavelengths such as radio does not yield remnants that are confidently classified as Type Ia remnants. Determining the remnant type is not easy. For large remnants, which have swept up several hundreds, even thousands, of solar masses \citep{Green84}, the abundances are determined by the surrounding material, and it is not possible to easily separate the ejecta and determine its composition. Although the presence of a central compact object can distinctly define a core-collapse SNR, the lack of one is not enough to confidently classify a remnant as Type Ia, especially when remnants become old and diffuse. It is possible that the lack of remnants may perhaps be due to misclassification, but that begs the question of why the ones in the LMC are better classified. At the low temperatures prevalent in these large remnants, the high level of absorption by the galactic disk makes it harder to observe spectral lines of elements such as O, Ne, Mg, and Fe, emitted at energies $kT<2$ keV. The low ISM density between the LMC and us allows us to discern these lines better than for the Galactic SNRs \citep{Maggi16}, which may be one reason why the LMC ones can be better classified as Ia's. It seems clear though that there is a deficit of large remnants in the Galaxy, both Type Ia and core-collapse.

\subsection{Core-Collapse Remnants}

The blue solid colored region in Figure \ref{fig:CC} outlines the range of detectability of objects by modern-day X-ray satellites, at the distance of the LMC. This is highly approximate, and the limit is taken to be at the lowest luminosity remnant observed in our sample, G38.7-1.4's (luminosity $7.29\times 10^{33}$ erg s$^{-1}$). The LMC population has a similar observed-luminosity lower bound of $7\times10^{33}$ erg s$^{-1}$ \citep{Maggi16}.
When the X-ray luminosity of the remnant exceeds this value, for a given density, it can be assumed to be detectable with current instruments. The red lines in Figure \ref{fig:CC} show how the luminosity of a remnant evolving in a given density would vary with increasing diameter, used as a proxy for time. These lines assume propagation into a uniform medium, with an initial explosion energy of $10^{51}$ erg. A smaller explosion energy would move the lines to the left, and a larger one to the right. We use the temperature given by equation (\ref{eqn:st_shock_temp}) to get the X-ray luminosity from equation (\ref{eqn:lumx}).

\subsubsection{Small Size Core-Collapse Remnants in the LMC}  Examination of the figure reveals a lack of small ($D$ $<$ 10pc) LMC CC SNRs. In the Milky Way, five such remnants are seen in X-rays, with four of them being relatively bright compared to the galactic population ($L_x > 2 \times 10^{36}$ erg s$^{-1}$). It is clear that small remnants in the LMC, evolving in a density of around 0.1 cm$^{-3}$ would not be detectable, whereas those evolving in a medium with density $> 5$ cm$^{-3}$ would generally be detectable. Remnants evolving in a medium with a number density around 1 cm$^{-3}$ would only be detectable with a diameter of 4 pc or larger. While these numbers are approximate, they provide a feel for the detectability, and show that remnants evolving in a low density would not be detected even with modern-day telescopes. A low density medium can be formed by stellar winds from a massive star, which evacuate the surrounding medium to form a wind-blown bubble surrounded by a dense shell \citep{weaveretal77, Dwarkadas05}. The density within the shocked wind medium, which occupies a large volume of the bubble, could be lower than 0.1 cm$^{-3}$. If the subsequent SN evolves in this low-density medium, it is unlikely that the remnant will be visible until it impacts with the dense shell that forms the boundary of the wind bubble. The interaction with the dense shell would result in an increase in the X-ray emission \citep{Dwarkadas05}, making the remnant X-ray bright.

\begin{figure*}
    \includegraphics[width=\textwidth,angle=0]{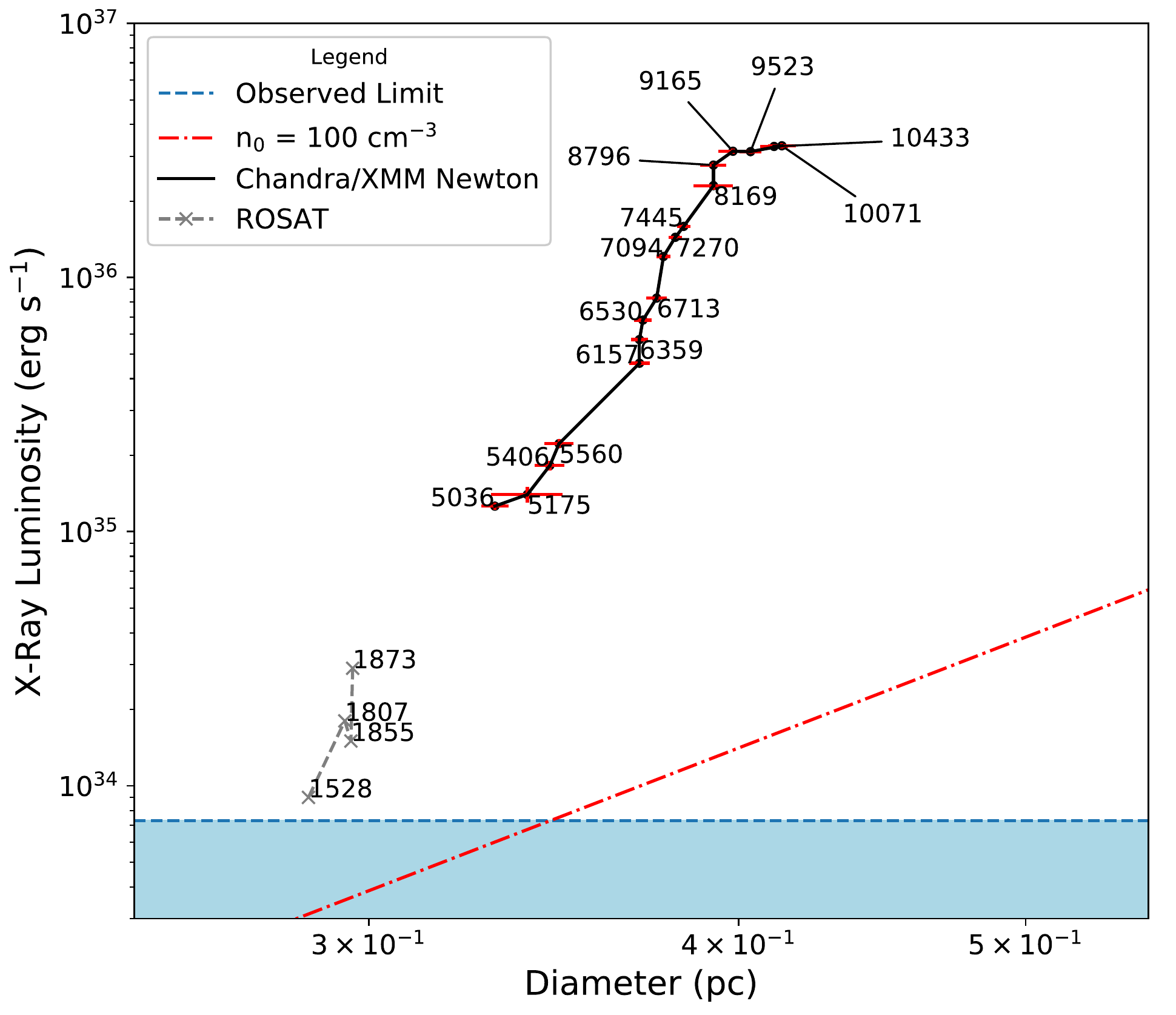}
    \caption{The evolution of LMC remnant SN1987A in the diameter-$L_x$ plane with age given in days. Observations from ROSAT in \citet{Schlegel95} are shown in gray. Chandra and XMM Newton Observations are downloaded from the Supernova X-Ray Database \citep{RossDwarkadas17}, and are mainly from \citet{Frank16}.
    \label{fig:SN1987A}}
    
\end{figure*}

Such a variation in X-ray luminosity has indeed been seen in SN 1987A (Figure \ref{fig:SN1987A}). The X-ray emission from the SN was at very low levels in the first two years or so, and almost undetectable, as the SN shock was expanding within the low density progenitor wind. Around an age of 3 years it began to interact with a dense region of ionized material  \citep{cd95}, causing the X-ray emission to rise. The rise in X-ray emission continued as the shock expanded within this ionized HII region, followed by an interaction with the dense shell (equatorial ring) surrounding the SN. This shell is theorized to have been formed by the fast wind from the blue-supergiant progenitor interacting with, and sweeping up, a prior slow wind from a red-supergiant phase \citep{lm91, bl93,deweyetal12}. Observations show that the shock has emerged from the dense shell around 10,000 days after the explosion \citep{franssonetal15}. The X-ray emission has levelled off and begun to slowly decrease. SN 1987A is an exception in that it is a very small remnant (diameter 0.4 pc) in the LMC that still happens to be visible because of the early interaction with the dense wind medium carved out by the progenitor blue-supergiant star.

The fact that no other SNR with a small size is seen indicates that SN 1987A is rare, as already suggested by its unusual blue supergiant progenitor. It is though surprising that not a single other small core-collapse SNR is seen in the LMC, although Ia's are. Part of the explanation may lie in the fact that, as pointed out earlier, the X-ray luminosity from a Type Ia remnant expanding in a constant density will increase with time, whereas that from a core-collapse remnant expanding in the wind of its progenitor star will decrease with time. Many core-collapse remnants are seen with a size between 10-20 pc, with luminosities ranging from 10$^{34}$ to a few times 10$^{37}$ erg s$^{-1}$. The inference is that perhaps the younger CC SNRs are expanding in a lower density medium. Extending this further we may infer that the younger CC remnants are evolving in wind bubbles that extend to about 10 pc and only emerge from the bubbles after an extended period. 

The variation of wind parameters with metallicity is not very clear. The wind velocity in massive stars was found to vary with metallicity Z as v$_{\infty} \propto Z^{0.13}$ (\citet{Leitherer92}). Mass-loss rates for stars with T$_{eff} \ge 25,000$ K varied as $\dot{M} \propto Z^{0.69}$, while those for lower temperature B supergiants varied as $\dot{M} \propto Z^{0.64}$ \citep{vinketal01}.  A more recent paper \citep{vs21} suggests that for cooler B supergiants, the wind terminal velocity is independent of metallicity, while mass-loss rates vary as $\dot{M} \propto Z^{0.85}$. For hotter O stars, the wind terminal velocity varies as v$_{\infty} \propto Z^{0.19}$, whereas the mass-loss rate variation is shallower, going as $\dot{M} \propto Z^{0.42}$.  In general, it is clear that the wind density ($\propto \dot{M}/\text{v}_{\infty}$) near the star will decrease at lower metallicty. The LMC metallicity is about a factor of 3 lower than that of the Milky Way \citep{Wilms00,Maggi16}. The stellar wind density in the LMC would therefore be up to a factor of 3$^{0.85} = 2.55$ lower than that in our Galaxy, and consequently the X-ray luminosity could be as much as a factor of 6.5 lower than in the Milky Way. While not a huge factor, it does contribute to the emission from a SNR within a wind-blown bubble falling below the detection threshold.

\subsubsection{Density Variations with Time} 

Many core-collapse remnants in Figure \ref{fig:CC} appear to lie at a position that suggests a density inconsistent with the measured density around the SNR. For example, the ambient density around the SNR Cas A is estimated to be 1 cm$^{-3}$ at the present time by  \citet{leeetal14}, and up to a factor of 2 higher by \citet{hl12}. However, its position in Figure \ref{fig:CC} indicates a density greater than 10 cm$^{-3}$.  While there is some ambiguity in the red lines due to the application of S-T temperature estimates to younger SNRs, it is clear that its luminosity indicates a higher density than is currently found. These lines are drawn assuming that the remnant is evolving in a constant density. If the density is not constant, but continuously evolving, its position on the plot would reflect an average density. In Cas A's case, it is evolving in a wind. For a steady wind, the density decreases with time as r$^{-2}$. The density in which Cas A is evolving was therefore much higher in the past, and the average density suggested by its position would be overall higher than the current density around the remnant. However, this may not be enough to account for the large luminosity. The average density of a wind with constant parameters, and therefore an r$^{-2}$ decline, can be easily shown to be 3 times the current density ahead of the shock. Therefore, the average wind density would be the equivalent of a density of 3-6 cm$^{-3}$, still lower than what its position suggests. It is possible that the wind parameters were not constant, leading to a higher average density.  Another factor, similar to Tycho's SNR, is the high luminosity of the shocked ejecta in Cas A. Although the ejecta mass may be small compared to the swept-up mass, the ejecta contribution to the X-ray luminosity is large, as shown by \citet{hl12}, resulting in a high overall luminosity compared to its current position.

Other SNRs may also have evolved similarly to Cas A, in a red supergiant wind, and are located not too far away from Cas A in Figure \ref{fig:CC}. \citet{Koo07} found $H_2$ and [Fe II] filaments within the radio shell of the SNR G11.2-0.03. Their results led them to suggest that the remnant was a Type IIL/b interacting with a red supergiant wind. \citet{BorkowskiReynolds17} found no evidence of a wind-blown bubble in Kes 73, and suggested that it was a Type IIP remnant that had expanded in a red supergiant wind. In each of these cases, a higher ambient density suggested by the red lines compared to that deduced for their current position suggests that the density of the surrounding medium was higher earlier, compatible with wind evolution, perhaps coupled with other factors as in Cas A.

In other cases the opposite may be true, i.e. given its current position, the density indicated by the plot is lower than what is currently measured. The classic case is of SN 1987A. As seen in Figure \ref{fig:SN1987A}, the SN luminosity increased by almost a factor of 100 for a very small increase in size of 0.1 pc or so. The current density in the plot is smaller than the density of 10$^4$ cm$^{-3}$ measured for the dense ring with which the remnant was interacting \citep{mf16}. This clearly indicates that the density was much lower in the past, as we know from its evolution in the wind-blown bubble.

\citet{Gaensler08} suggested that SNR G350.1-0.3 was interacting with a molecular cloud or a very dense wind. They provided two estimates of ambient density, giving a range of 25-600 cm$^{-3}$, which we adjust in table \ref{table:Gal} to reflect the distance uncertainty. The high end of this range is derived from the assumption that G350.1-0.3 is in the S-T phase, a reasonable assumption given its position on the plot. However, its position in Figure \ref{fig:CC} suggests a density closer to 10 cm$^{-3}$, lower than the proposed range. This may indicate that the density was lower earlier, and is now higher. It suggests evolution in a wind-blown bubble at an earlier stage, or in a constant density medium with density similar to the average ISM density.

\subsection{Comparing Distance Estimates} \label{subsec:compare}

\begin{figure*}
    \includegraphics[width=\textwidth,angle=0]{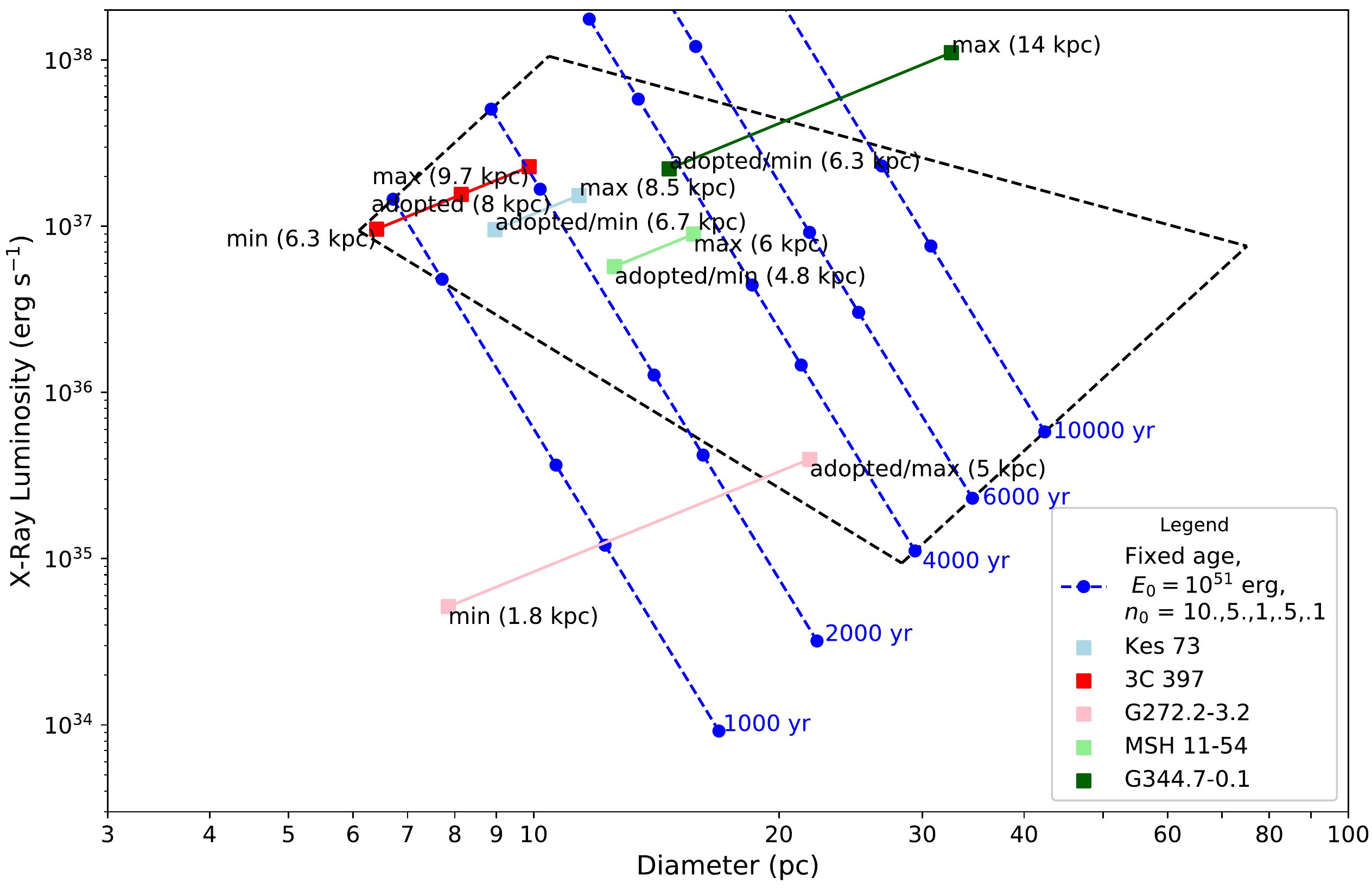}
    \caption{Plotting the distance range on selected SNRs. Reference lines for constant density S-T stage SNRs are provided to evaluate how well a given distance estimate agrees with the model. The S-T region is plotted under the assumption of a constant ambient density. Note that with alternative distance estimates, the resulting $n_0$ and age may differ from the value ranges given in Table \ref{table:Gal}.}
    \label{fig:compare}
    
\end{figure*}

The large error bars on galactic distance measurements can have large effects on the calculated luminosity and size of a SNR. Figure \ref{fig:compare} shows the change in the remnant's properties when an alternative distance value is used, and how such an alternate distance could be deduced from the plot. The values labeled ``adopted" are those using which the size and luminosity in Table \ref{table:Gal} are calculated. The remnants G344.7-0.1 and G272.2-3.2 had prior distance estimates that placed them away from the rest of the population and outside the S-T region. The outliers would have indicated extreme properties, and therefore the need for further investigation and refinement. The figure illustrates how the uncertainty in distance measurement can affect a remnant's size and luminosity, as shown for Kes 73, 3C 397, and MSH 11-54. 

The distance to G344.7-0.1 was first calculated by \citet{Dubner93} using the $\Sigma$-$D$ relation to be 14 kpc. \citet{Combi10} adopted this distance and derived the properties accordingly. A distance of 14 kpc would make G344.7-0.1 the largest and most luminous SNR in the galactic population. However, the $\Sigma$-$D$ relation is known to be an unreliable distance indicator \citep{Green84}. \citet{Giacani11} revised the distance to approximately 6.3 kpc by studying HI absorption and emission. As seen in the figure, this places it in the S-T region and makes its properties not seem as extreme. 

In retrospect, an inspection of the figure would have suggested a need to revise the distance or postulate exceptional properties for this SNR. G344.7-0.1 has a shell-type morphology with significantly increased brightness in the western region, interpreted by \citet{Combi10} as interaction with a dense molecular cloud. \citet{Giacani11} corroborated the proposed cloud interaction with observations of a neutral hydrogen region bordering the northern and western sides of G344.7-0.1's shell.
The east-west density gradient arising from the cloud interaction suggested that a constant density model was not applicable. The nearly spherical form of the shell implied that the shock had only recently encountered the cloud, otherwise it would have created an indentation on the western region where the shock was slowed down by the denser medium.
\citet{Combi10} derived an age of 6000 yrs by dividing the upper limit of the ionization timescale by the electron density calculated from the emission measure. For a diameter around 30 pc, this implied that G344.7-0.1 previously evolved in a low density medium of $n_0 < 0.1$.
This density would be lower than estimates for other galactic Type Ias, although not completely unreasonable. Using the age of 3000 yrs found by \citet{Giacani11} at a distance of 6.3 kpc, the pre-cloud $n_0$ is more typical of an average ISM density of 1 cm$^{-3}$.
This example demonstrates how nonconformity in the size to luminosity relation can signal that a distance estimate has room for improvement. In G344.7-0.1's case, it appears to indicate the smaller distance of 6.3 kpc.

Similarly, refinement of the distance to G272.2-3.2 can be visualized on the plot. ROSAT observations by \citet{Greiner94} yielded a distance of d$\approx$1.8 kpc using the hydrogen column density to extinction relationship. This value placed G272.2-3.2 away from all other SNRs on the plot, giving an age range of 400-4000 yr \citep{Greiner94}. Other Type Ias of similar size at a similar distance tend to be much more luminous. This can be contrasted with the larger distance of d$\approx$5 kpc derived from the color excess to column density relationship of stars in the direction of G272.2-3.2 \citep{Harrus01}. At this distance, its properties appear to be more in agreement with other Type Ia SNRs.

\subsection{Excluded Remnants}
Of the over 380 detected SNRs in the Milky Way \citep{FerrandSafi-Harb12} and the 92 detected SNRs and SNR candidates in the LMC \citep{Bozzetto17,Maggi16,Yew21}, we have considered a subset of 34 Galactic and 59 LMC remnants, for which data were available, in our X-ray study. It is important to verify that our sample population is representative of the whole. For the Milky Way, our requirement of obtaining the X-ray flux over the entire remnant excludes more than half of the SNRs that exhibit thermal X-ray emission. 

\begin{figure*}
    \centering
    \includegraphics[width=\textwidth]{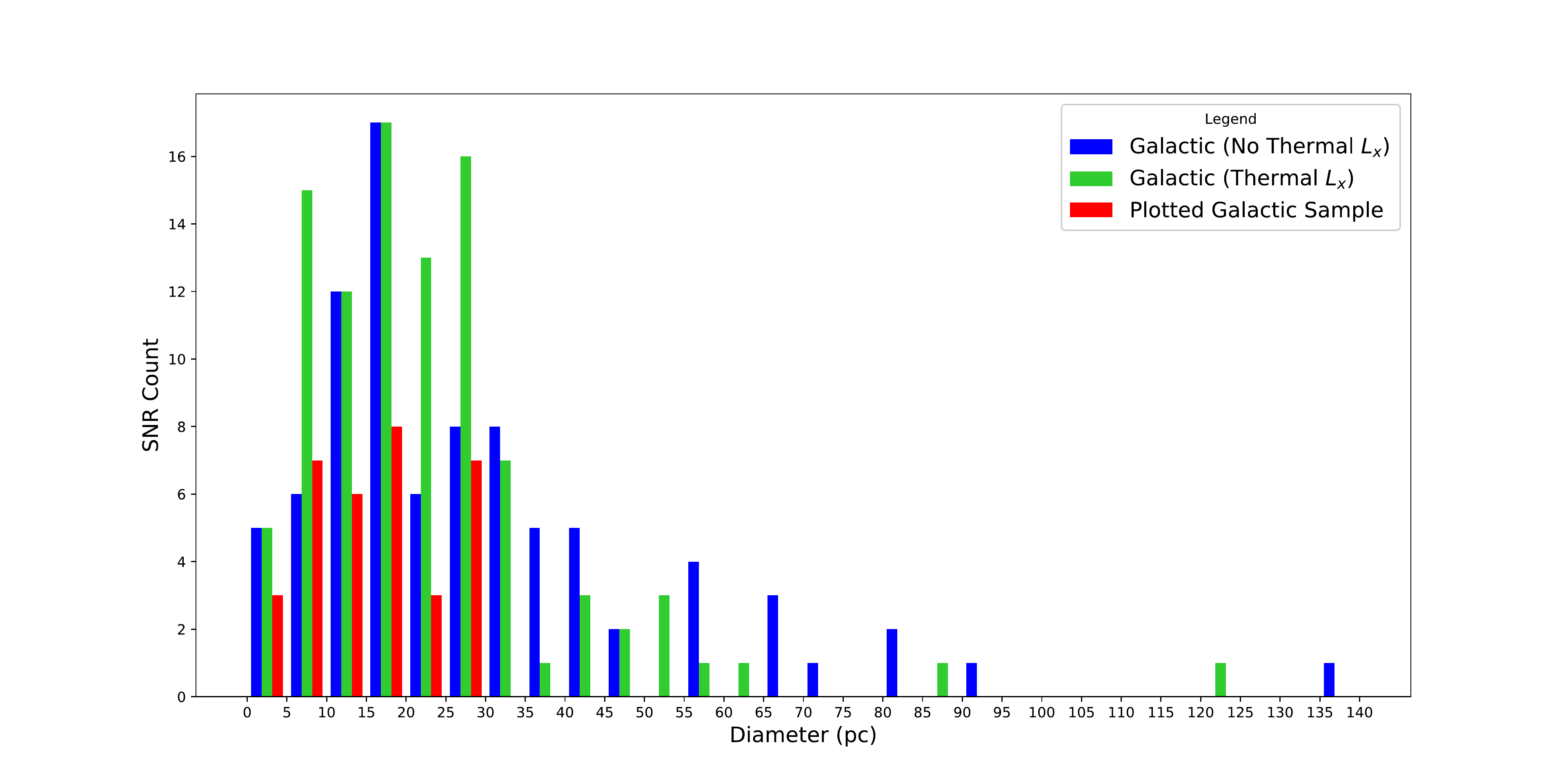}
    \includegraphics[width=\textwidth]{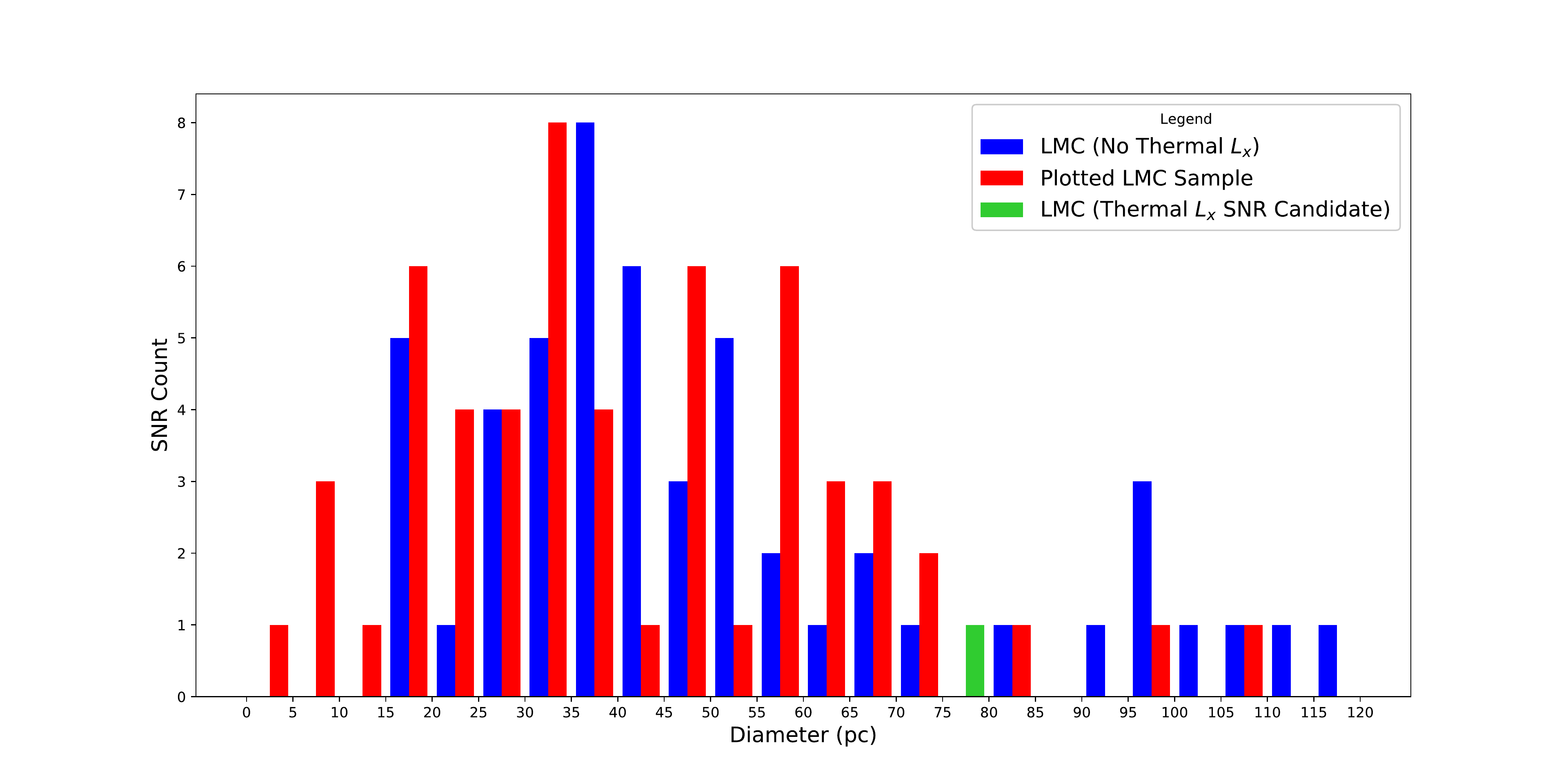}
    \caption{\textbf{Top:} Histogram of all galactic SNRs with a distance estimate cataloged by \citet{FerrandSafi-Harb12}. Our sample for fully imaged, thermally emitting SNRs (red) are a subset of the entire thermally emitting SNRs group (green). All SNRs without significant thermal X-rays are combined into the blue group. \textbf{Bottom:} Histogram of combined population of known SNRs and SNR candidates in the LMC recorded by \citet{Maggi16,Bozzetto17,Yew21}. The thermal X-ray emitting LMC SNRs (red) are already consistent with the remnants we use, with only one SNR candidate having confirmed thermal X-ray emission.}
    \label{fig:excluded}
\end{figure*}

The total Galactic thermal X-ray emitting SNR population is shown in Figure \ref{fig:excluded}. The overall distribution of the entire set is similar to the subset used in our study, except for higher counts at smaller diameters, and a slowly decaying tail at larger diameters. The lack of SNRs at larger diameters in our sample can be attributed to several factors. First, the total galactic population has a falloff at large sizes so there are fewer remnants. Furthermore, large X-Ray SNRs might not be fully imaged due to their large angular size, that exceeds the field of view on Chandra and perhaps XMM-Newton. We do not include galactic remnants that have not been typed, which is difficult for large, diffuse remnants without a compact central object. The cutoff in our sample occurs around diameters of 30 pc, which agrees with the steep falloff of the entire thermal X-ray population. While the lack of large Type Ias in our sample can be attributed to the difficulty of typing old galactic remnants, the lack of CC remnants with diameters over 30 pc in our sample is presumably due to incomplete imaging, as indicated by the presence of many Galactic SNRs that are dominated by thermal X-ray emission. We conclude that despite the relatively small sample size used herein, it is a reasonable representation of the total galactic population. 

The lack of small, young CC X-ray remnants in the LMC is also reflected in optical and radio data. All LMC remnants with a diameter less than 15 pc are included in our sample. This implies that besides SN1987A, no other small CC remnants have been detected, supporting the assertion that CC SNRs are evolving in the progenitor's wind-blown bubble. Interestingly, small SNRs in the Milky Way are not necessarily dominated by thermal X-rays, while in the LMC all the small SNRs observed show thermal emission. 

\begin{figure*}
    \centering
    \includegraphics[width=\textwidth]{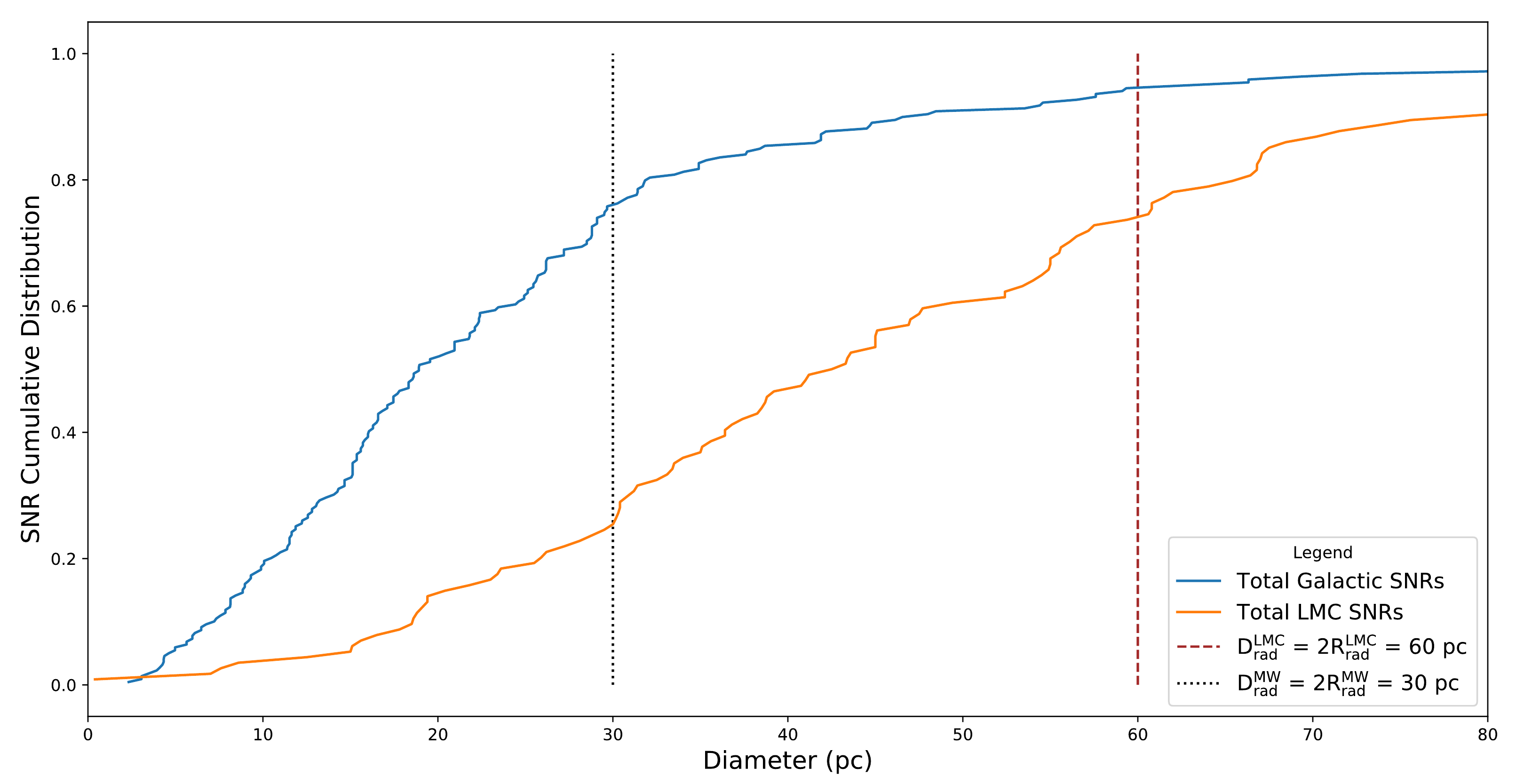}
    \caption{Total observed SNR count rates with respect to diameter. The LMC and Milky Way S-T cutoff radii are marked in dashed brown and dotted black, respectively.}
    \label{fig:cumulative}
\end{figure*}

\subsection{Cumulative Distribution of Remnants}
\label{subsec:total}

The cumulative distribution of SNRs in each galaxy is given in Figure \ref{fig:cumulative}. The total SNR cumulative curves, including all currently untyped remnants, are used in our analysis.  The CC cumulative distribution curve in each galaxy agrees well with the total SNR curve. The Type Ia numbers are smaller, but the agreement of the total and CC distribution curves suggests that Type 1a SNe follow a similar distribution. The cumulative distribution curves are assumed to level off when the SNRs become radiative, since radiative remnants are expected to fade quickly and become undetectable. Combined with the lack of young remnants, this suggests the majority of SNRs that are detected must be in the S-T phase \citep{Badenes2010}.

The radius at which a SNR enters the radiative phase \citep{Truelove&Mckee99} is given by equation (\ref{eqn:Rrad}),
where the metallicity correction factor $\zeta_m$ is proportional to the metallicity of the surrounding environment \citep{Badenes2010}. We can then compare the size at which Milky Way and LMC remnants appear to have entered the radiative phase. Assuming a constant SNR creation rate implies that the count rate should have a power law relation with radius \citep{Badenes2010}. Each evolutionary phase of a SNR has a power-law relation between radius and time, with the later stages having a smaller power-law index as the shock is continually slowing down. The radius-time power-law index can be close to 1 in the ejecta dominated stage, but will go to 0.4 by the S-T stage. Thus the cumulative distribution curve should have a derivative that changes with evolutionary state. While the slow expansion in the radiative phase should lead to higher concentrations of remnants at large diameters, we do not see an increasing count rate at large diameters, likely due to old remnants becoming harder to detect as they lose energy and merge into the ISM. We denote as $R_{\text{rad}}^{\text{MW}}$ and $R_{\text{rad}}^{\text{LMC}}$ the radius at which remnants in the Galaxy and LMC enter the radiative phase, or alternatively the maximum radius in the S-T stage. Since we expect a similar count rate throughout the S-T phase, we choose $R_{\text{rad}}^{\text{MW}}$ and $R_{\text{rad}}^{\text{LMC}}$ such that they are the threshold values from Figure \ref{fig:cumulative} marking the end of the approximately linear count rate seen once the S-T phase is entered. Taking $R_{\text{rad}}^{\text{MW}} = 15$ pc and $R_{\text{rad}}^{\text{LMC}} = 30$ pc, we get a ratio of their average ambient densities of
\begin{align}
    \frac{n_0^{\text{LMC}}}{n_0^{\text{MW}}} &= \left(\frac{R_{\text{rad}}^{\text{MW}}}{R_{\text{rad}}^{\text{LMC}}} \left(\frac{Z_{\text{LMC}}}{Z_{\text{MW}}}\right)^{-1/7} \right)^{7/3} \\
    \frac{n_0^{\text{LMC}}}{n_0^{\text{MW}}} & = \left(\frac{1}{2} \left(\frac{1}{3} \right)^{-1/7} \right)^{7/3} \\
    & \approx 0.3,
\end{align}
when taking the metallicity $Z$ of the Milky Way to be a factor of 3 higher than in the LMC \citep{Wilms00,Maggi16}. 
Thus the difference between the cumulative distributions can be explained as a difference in density. \citet{Maggi19} provide an estimate for a typical density in SNR forming regions of the LMC and a median value of $n_0^{\text{LMC}} = 0.1$ cm$^{-3}$ is found. This can be seen in Figure \ref{fig:CC} by extending the $n_0$=0.1 line through the cluster of large LMC remnants. Given the density ranges in the Galactic data set, we find that the median ambient density around Galactic SNRs is 0.5-1.0 cm$^3$.

To analytically describe the S-T regions of the cumulative curves, a power law relationship of the form 
\begin{equation}
\label{eqn:mle}
\frac{dN}{dD} \propto D^{\alpha}
\end{equation}
\noindent is assumed. An estimate for the $\alpha$ value that produces a probability density most likely to result in the observed data is obtained by assuming that the SNR size distribution is a Poisson process and constructing a maximum likelihood estimator (MLE). A complete derivation is given in \citet{Badenes2010}. Following their equations 1-5, we find MLEs for $\alpha$ in the Milky Way and LMC to be: 
$\hat{\alpha}_{\text{MW}} = 0.33 \pm 0.10$, $\hat{\alpha}_{\text{LMC}} = 0.47 \pm 0.16$
when considering remnants up to size $R_{\text{rad}}^{\text{MW}} = 15$ pc in the Milky Way and $R_{\text{rad}}^{\text{LMC}} = 30$ pc in the LMC. 

\citet{Badenes2010} and \citet{Vink20} derive the predicted SNR count rate in the S-T phase. Assuming the simplest case of all remnants expanding in a medium with the same density, and occurring at a constant creation rate, the derivative of the count rate scales as $\frac{dN}{dD} \propto D^{3/2}$, where D is the diameter. Since it is clearly not the case that all the SNRs evolve within the same density medium, we investigate what kind of probability distribution of ambient densities could exist in each galaxy to make the model fit the results, if the density in which each SNR was expanding is drawn from this distribution.  The ambient density probability distribution $P$ within each galaxy is assumed to follow a power law
\begin{equation}
\centering
\frac{dP}{dn_0} \propto n_0^{\beta}.
\end{equation}
Solving equation ($\ref{eqn:Rrad}$) for $n_0$ gives us the density that produces the max S-T radii:
\begin{equation}
n_0 \propto R^{\delta},
\end{equation}
where $\delta$ is between -7/3 and -5/2 and the variation comes from the temperature dependence of $R_{\text{rad}}$ \citep{Badenes2010}.
This correction for the density distribution $\frac{dP}{dn_0}$ gives \citep{Badenes2010}
\begin{equation}
\label{eqn:prob}
\frac{dN}{dD} \propto D^{\delta(\beta+3/2)+3/2}.
\end{equation}

Equating the exponents of Equation \ref{eqn:mle} and Equation \ref{eqn:prob} using the estimated MLEs for $\alpha$, we find that $-0.96 < \beta_{\text{MW}} < -1.07$ and $-0.99 < \beta_{\text{LMC}} < -1.15$. We conclude that a density distribution described by a power law with index $\beta \approx -1$ is sufficient to explain the size dependence of the SNR count in the S-T phase in both the Milky Way and LMC. Fixing the size dependence of the distribution in the S-T phase and fixing a point of the cumulative count at a given size is enough to characterize the $N$ vs $D$ distribution curve. Since each population's count has a similar $D$ dependence, the discrepancy between the curves can be attributed to the ratio of the average densities of the regions surrounding the SNRs. The high frequency of LMC remnants at large diameters suggests that this density difference is more likely due to the ambient ISM than stellar winds. 

Since ejecta-dominated LMC core-collapse remnants may possibly fall within an undetectable X-ray luminosity range due to evolution in a wind-blown bubble, as opposed to being absent altogether from the population, the true cumulative distribution curve for the LMC may grow faster at small diameters. This would increase the cumulative percentage prior to the S-T phase, but the power law index of the cumulative curve in the S-T phase would remain the same. The above analysis relies on S-T remnants, so our results are robust, assuming that our sample of S-T remnants is representative of the whole.

\section{Discussion and Conclusions}
\label{sec:concl}

In this paper we have studied the evolution of SNRs over their lifetimes. We use the X-ray emission from the SNR as a guide to its evolutionary phase, combined with the size of the remnant as a proxy for its age. We find that the majority of observed remnants lie within the Sedov-Taylor phase as expected, and their properties are in general agreement with theoretical models. Remnants that deviate from the models include those whose luminosity arises mainly from non-thermal emission, the large majority of which we have not taken into account; those whose distance was not properly estimated, leading to errors in their properties; those where the ejecta luminosity was high in proportion to the ejected mass; and those which were evolving in a medium whose density changed with time. The latter would include remnants which initially expand within a stellar wind whose density is decreasing with time, and then a constant density interstellar medium; remnants which expand within a low density wind bubble before interacting with a higher density shell; as well as those initially expanding within the ISM and then impacting high density clouds.

Type Ia SNRs in the Milky Way and LMC  appear to evolve in a manner consistent with the S-T model for evolution in a constant density medium, within the Sedov-Taylor boundaries given by \citet{Dwarkadas&Chev98} and \citet{Truelove&Mckee99}. The Milky Way's thermal X-ray Type Ia SNRs agree generally well with theoretical S-T stage predictions. Many Type Ias in the LMC fall outside our S-T predictions due to a low luminosity and large size. This could be because they are indeed S-T phase SNRs with $n_0 < 0.1$, or because they have entered, or are well-within, the radiative phase. The relatively low density of the LMC compared to the Milky Way shown by \citet{Maggi19} and in section \ref{subsec:total} suggests that some of them could be S-T remnants expanding within a very low density. Remnants in the Milky Way that have evolved beyond the S-T phase are underrepresented in the Galactic X-ray population. This is true for both Type Ia and CC remnants. This is presumably in part due to our selection constraints, because galactic remnants with large angular sizes are less likely to be imaged in full, assuming that they are hot enough to be detected in X-rays. Old, large SNRs without a detected central object are also more difficult to type. The clustering of SNRs near our detection limit, as well as near the XMM-Newton detection limit used by \citet{Maggi16} and \citet{Ou18}, shows that our current catalog of thermally emitting X-ray SNRs in the Milky Way and Magellanic clouds would greatly benefit from advancements in instrument sensitivity.

The LMC population is lacking in small, young core-collapse remnants when compared to the Milky Way. Besides SN1987A, the next smallest CC X-ray SNR detected in the LMC is J0540-6920 (SNR 0540-69.3) with a diameter of 15.6 pc \citep{Bozzetto17}. Young CC remnants may first expand into a progenitor-blown wind bubble, where high speed winds can sweep up the ambient ISM to create a low density environment \citep{Dwarkadas05} that renders young SNRs undetectable. Also, the lower metallicity of the LMC leads to a decrease in stellar mass loss rates \citep{vinketal01,vs21}, further decreasing the density into which the young CC remnants are expanding. SN1987A has a high luminosity because the SN shock encountered the high density shell bordering the wind-blown bubble, considerably increasing its luminosity. Sudden emergence from a bubble into its dense swept-up edge may also explain the bifurcation in LMC CC remnants between 10-30 pc observed by \citet{Ou18}. Adding in the Milky Way's CC SNRs, we get a more uniform distribution in this range on Figure \ref{fig:CC}, while also observing that the more X-ray luminous Galactic CCs tend to be the smallest ones. The large vertical spread of these mid-sized Galactic and LMC SNRs could be from remnants increasing in luminosity from the undetectable regime as the SN shock interacts with the dense edge of a bubble, similar to SN1987A (Figure \ref{fig:SN1987A}). This population could also include SNRs evolving in smaller wind bubbles, which have already emerged from the bubble and are now interacting with a lower-density medium compared to the dense shell of the bubble. SN 1987A will fall into this category soon.

The Galactic X-ray SNR population includes well resolved supernova remnants whose internal features and surrounding medium can be well studied. The size and X-ray luminosity of Galactic SNRs suffers from the relatively large uncertainty in their distance measurements compared to SNRs in nearby galaxies whose distance is well known. However, theoretical models, combined with SNRs whose distances are accurately known, can be used to improve the parameters for remnants with imprecise distances. Given a new distance estimate, our plots provide a quick method of estimating its compatibility with the remainder of the observed population. Examples of this process are given for G272.72-3.2 \citep{Greiner94, McEntaffer13} and G344.7-0.1 \citep{Combi10,Giacani11}, where our plots suggest either that the initial distances were incorrect, or the remnants had extreme physical properties. Subsequent investigations showed that the distance estimate could be improved, bringing the properties in line with the rest of the SNR population. 

These estimates are more applicable to Type Ia remnants, because of the complexity of the wind-blown, possibly multi-layered, bubbles that core-collapse SNRs interact with during the ED phase. G308.3-1.4 and G330.2+1.0 are two CC SNRs located near each other on the diameter-vs-$L_x$ plot and currently interacting with similar ambient densities, but G308.3-1.4 (G308.4-1.4) is estimated to be 4000-6500 years older than G330.2+1.0 \citep{Borkowski18,PrinzBecker12}. The $\Sigma-D$ relation applied to G308.3-1.4 \citep{DeHorta13} gives a distance of $19 \pm 11$ kpc that yields a diameter more suited to its age, but the large uncertainties allow for overlap with \citet{PrinzBecker12}'s distance estimate of 9.8$^{+0.9}_{-0.7}$ kpc. While this illustrates the limitations of evaluating CC SNR distances with constant ISM models, the plotted proximity of these two SNRs, combined with the age estimates, indicates that G308.3-1.4 previously expanded in a far denser ISM than what its shock is currently encountering.

While the statistics are small, this survey of Galactic and LMC X-ray remnants reveals a lack of young remnants still expanding in the dense winds of their progenitor stars. Our Galaxy has one young remnant, Cas A, continuing to expand in the wind of its progenitor at an age of about 340 years.  Other SNRs that may have evolved in a RSG wind include G11.2-0.03 and Kes 73. \citet{BorkowskiReynolds17} conclude that Kes 73 is now interacting with the ambient ISM, so it is difficult to determine how long it was interacting with the freely expanding wind. G11.2-0.03 is supposedly much older than Cas A, and while it was presumably interacting with a wind at some point, the observations are not decisive on whether the interaction is still continuing. The LMC appears to have no remnants still expanding in stellar winds. This may suggest that dense, extended winds like that of the remnant Cas A are rare. 

Young core-collapse SNe are visible for several decades in X-rays \citep{dg12, vvd14, drrb16, RossDwarkadas17, bocheneketal18}. Their luminosity is typically decreasing with time, suggesting expansion in a wind. The densities and mass-loss rates inferred for these winds can be quite high, with mass-loss rate $ \dot{M} > 10^{-3} M_{\odot}$ yr$^{-1}$ for the highest luminosity Type IIn SNe \citep{franssonetal14, chandraetal15, drrb16}. Thus there is no doubt that dense winds exist around young SNe. In the IIn SNe however, after a few years the X-ray luminosity  generally falls off much faster than in an r$^{-2}$ wind \citep{dg12}. The oldest detected X-ray SNe \citep{rd20}, which are a few decades old, have X-ray luminosity comparable to, or lower than that of Cas A, which is almost 300 years older. If these old SNe continue to expand in a wind medium, their luminosity will continue to decrease, and will fall much below that of Cas A at the same age. Thus it does not appear that a dense extended wind is at all common. The fact that there are few known remnants within about 100-300 years, whereas many older ones are known, is consistent with the fact that an extended dense wind like that around Cas A is uncommon. Observations of the light echo from Cas A have indicated that it was a Type IIb SN, with a spectrum similar to that of SN 1993J \citep{krauseetal08}. SN 1993J is known to be a binary, and the signatures of a companion star have been inferred \citep{maundetal04, foxetal14}. It is probable that Cas A also had a binary companion, resulting in an extended dense wind. However, while the initial mass-loss rates of the wind in SN 1993J  may have been similar to that derived for Cas A, observations show that the X-ray emission from SN 1993J started decreasing at a much faster rate after about 3000 days \citep{dbbb14}, and its X-ray luminosity after 20 years was already lower than that of Cas A. It is clear that the conditions that lead to such extended mass-loss as in Cas A are not common.

A study of the distribution of all detected SNRs in the Milky Way and LMC suggests that densities in SNR forming regions of the LMC are on average about a factor of 3 lower than in the Milky Way. This should make LMC luminosities a factor of 9 lower than Galactic remnants at a given size. This can be seen in the Type Ia population for the diameters where the populations overlap. A large portion of the LMC population has diameters larger than any Galactic remnant in our population, which may also be due in part to evolution in a lower density medium. Lower average densities in the LMC also contribute to the lack of observable small CC remnants. 

The survey of all detected Galactic SNRs serves as a check on the validity of our results, which are a subset of the total population. SNR count rates in the Milky Way for the total thermal X-ray population, and for the subset used herein, show similar distribution rates. All confirmed thermal X-ray SNRs in the LMC are included in this study, giving confidence that our subset represents the true population. Adding further remnants will increase the total count, but will probably not alter the fundamental results derived herein.

\section*{Acknowledgements}
We thank the referee, Martin Laming, for a very careful reading of the paper, and for several very helpful comments and suggestions that have greatly improved the manuscript. We thank Roger Chevalier for useful comments that helped to improve this manuscript. This work was supported by National Science Foundation grant 1911061 awarded to the University of Chicago (PI: Vikram Dwarkadas).

\section*{Data Availability} All data are incorporated into the article and its online supplementary material.

\bibliographystyle{mnras}
\bibliography{SNR-bibtex}{}

\begin{thebibliography}{}
\makeatletter
\relax
\def\mn@urlcharsother{\let\do\@makeother \do\$\do\&\do\#\do\^\do\_\do\%\do\~}
\def\mn@doi{\begingroup\mn@urlcharsother \@ifnextchar [ {\mn@doi@}
  {\mn@doi@[]}}
\def\mn@doi@[#1]#2{\def\@tempa{#1}\ifx\@tempa\@empty \href
  {http://dx.doi.org/#2} {doi:#2}\else \href {http://dx.doi.org/#2} {#1}\fi
  \endgroup}
\def\mn@eprint#1#2{\mn@eprint@#1:#2::\@nil}
\def\mn@eprint@arXiv#1{\href {http://arxiv.org/abs/#1} {{\tt arXiv:#1}}}
\def\mn@eprint@dblp#1{\href {http://dblp.uni-trier.de/rec/bibtex/#1.xml}
  {dblp:#1}}
\def\mn@eprint@#1:#2:#3:#4\@nil{\def\@tempa {#1}\def\@tempb {#2}\def\@tempc
  {#3}\ifx \@tempc \@empty \let \@tempc \@tempb \let \@tempb \@tempa \fi \ifx
  \@tempb \@empty \def\@tempb {arXiv}\fi \@ifundefined
  {mn@eprint@\@tempb}{\@tempb:\@tempc}{\expandafter \expandafter \csname
  mn@eprint@\@tempb\endcsname \expandafter{\@tempc}}}

\bibitem[\protect\citeauthoryear{{Acero}, {Ballet}  \& {Decourchelle}}{{Acero}
  et~al.}{2007}]{abd07}
{Acero} F.,  {Ballet} J.,   {Decourchelle} A.,  2007, \mn@doi [\aap]
  {10.1051/0004-6361:20077742}, \href
  {https://ui.adsabs.harvard.edu/abs/2007A&A...475..883A} {475, 883}

\bibitem[\protect\citeauthoryear{{Badenes}, {Maoz}  \& {Draine}}{{Badenes}
  et~al.}{2010}]{Badenes2010}
{Badenes} C.,  {Maoz} D.,   {Draine} B.~T.,  2010, \mn@doi [\mnras]
  {10.1111/j.1365-2966.2010.17023.x}, \href
  {https://ui.adsabs.harvard.edu/abs/2010MNRAS.407.1301B} {407, 1301}

\bibitem[\protect\citeauthoryear{{Becker}, {Prinz}, {Winkler}  \&
  {Petre}}{{Becker} et~al.}{2012}]{Becker12}
{Becker} W.,  {Prinz} T.,  {Winkler} P.~F.,   {Petre} R.,  2012, \mn@doi [\apj]
  {10.1088/0004-637X/755/2/141}, \href
  {https://ui.adsabs.harvard.edu/abs/2012ApJ...755..141B} {755, 141}

\bibitem[\protect\citeauthoryear{{Berezhko} \& {V{\"o}lk}}{{Berezhko} \&
  {V{\"o}lk}}{2004}]{Berezhko04}
{Berezhko} E.~G.,  {V{\"o}lk} H.~J.,  2004, \mn@doi [\aap]
  {10.1051/0004-6361:20041111}, \href
  {https://ui.adsabs.harvard.edu/abs/2004A&A...427..525B} {427, 525}

\bibitem[\protect\citeauthoryear{{Bianco} et~al.,}{{Bianco}
  et~al.}{2011}]{Bianco11}
{Bianco} F.~B.,  et~al., 2011, \mn@doi [\apj] {10.1088/0004-637X/741/1/20},
  \href {https://ui.adsabs.harvard.edu/abs/2011ApJ...741...20B} {741, 20}

\bibitem[\protect\citeauthoryear{{Blondin} \& {Lundqvist}}{{Blondin} \&
  {Lundqvist}}{1993}]{bl93}
{Blondin} J.~M.,  {Lundqvist} P.,  1993, \mn@doi [\apj] {10.1086/172366}, \href
  {https://ui.adsabs.harvard.edu/abs/1993ApJ...405..337B} {405, 337}

\bibitem[\protect\citeauthoryear{{Bochenek}, {Dwarkadas}, {Silverman}, {Fox},
  {Chevalier}, {Smith}  \& {Filippenko}}{{Bochenek}
  et~al.}{2018}]{bocheneketal18}
{Bochenek} C.~D.,  {Dwarkadas} V.~V.,  {Silverman} J.~M.,  {Fox} O.~D.,
  {Chevalier} R.~A.,  {Smith} N.,   {Filippenko} A.~V.,  2018, \mn@doi [\mnras]
  {10.1093/mnras/stx2029}, \href
  {https://ui.adsabs.harvard.edu/abs/2018MNRAS.473..336B} {473, 336}

\bibitem[\protect\citeauthoryear{{Borkowski} \& {Reynolds}}{{Borkowski} \&
  {Reynolds}}{2017}]{BorkowskiReynolds17}
{Borkowski} K.~J.,  {Reynolds} S.~P.,  2017, \mn@doi [\apj]
  {10.3847/1538-4357/aa830f}, \href
  {https://ui.adsabs.harvard.edu/abs/2017ApJ...846...13B} {846, 13}

\bibitem[\protect\citeauthoryear{{Borkowski}, {Reynolds}, {Green}, {Hwang},
  {Petre}, {Krishnamurthy}  \& {Willett}}{{Borkowski}
  et~al.}{2014}]{Borkowski14}
{Borkowski} K.~J.,  {Reynolds} S.~P.,  {Green} D.~A.,  {Hwang} U.,  {Petre} R.,
   {Krishnamurthy} K.,   {Willett} R.,  2014, \mn@doi [\apjl]
  {10.1088/2041-8205/790/2/L18}, \href
  {https://ui.adsabs.harvard.edu/abs/2014ApJ...790L..18B} {790, L18}

\bibitem[\protect\citeauthoryear{{Borkowski}, {Reynolds}, {Williams}  \&
  {Petre}}{{Borkowski} et~al.}{2018}]{Borkowski18}
{Borkowski} K.~J.,  {Reynolds} S.~P.,  {Williams} B.~J.,   {Petre} R.,  2018,
  \mn@doi [\apjl] {10.3847/2041-8213/aaedb5}, \href
  {https://ui.adsabs.harvard.edu/abs/2018ApJ...868L..21B} {868, L21}

\bibitem[\protect\citeauthoryear{{Borkowski}, {Miltich}  \&
  {Reynolds}}{{Borkowski} et~al.}{2020}]{Borkowski20}
{Borkowski} K.~J.,  {Miltich} W.,   {Reynolds} S.~P.,  2020, \mn@doi [\apjl]
  {10.3847/2041-8213/abcda7}, \href
  {https://ui.adsabs.harvard.edu/abs/2020ApJ...905L..19B} {905, L19}

\bibitem[\protect\citeauthoryear{{Bozzetto} et~al.,}{{Bozzetto}
  et~al.}{2017}]{Bozzetto17}
{Bozzetto} L.~M.,  et~al., 2017, \mn@doi [\apjs] {10.3847/1538-4365/aa653c},
  \href {https://ui.adsabs.harvard.edu/abs/2017ApJS..230....2B} {230, 2}

\bibitem[\protect\citeauthoryear{{Bruenn} et~al.,}{{Bruenn}
  et~al.}{2016}]{Bruenn16}
{Bruenn} S.~W.,  et~al., 2016, \mn@doi [\apj] {10.3847/0004-637X/818/2/123},
  \href {https://ui.adsabs.harvard.edu/abs/2016ApJ...818..123B} {818, 123}

\bibitem[\protect\citeauthoryear{{Burrows}, {Vartanyan}, {Dolence}, {Skinner}
  \& {Radice}}{{Burrows} et~al.}{2018}]{Burrows18}
{Burrows} A.,  {Vartanyan} D.,  {Dolence} J.~C.,  {Skinner} M.~A.,   {Radice}
  D.,  2018, \mn@doi [\ssr] {10.1007/s11214-017-0450-9}, \href
  {https://ui.adsabs.harvard.edu/abs/2018SSRv..214...33B} {214, 33}

\bibitem[\protect\citeauthoryear{{Carlton}, {Borkowski}, {Reynolds}, {Hwang},
  {Petre}, {Green}, {Krishnamurthy}  \& {Willett}}{{Carlton}
  et~al.}{2011}]{Carlton11}
{Carlton} A.~K.,  {Borkowski} K.~J.,  {Reynolds} S.~P.,  {Hwang} U.,  {Petre}
  R.,  {Green} D.~A.,  {Krishnamurthy} K.,   {Willett} R.,  2011, \mn@doi
  [\apjl] {10.1088/2041-8205/737/1/L22}, \href
  {https://ui.adsabs.harvard.edu/abs/2011ApJ...737L..22C} {737, L22}

\bibitem[\protect\citeauthoryear{{Caswell}, {Haynes}, {Milne}  \&
  {Wellington}}{{Caswell} et~al.}{1983}]{Caswell83}
{Caswell} J.~L.,  {Haynes} R.~F.,  {Milne} D.~K.,   {Wellington} K.~J.,  1983,
  \mn@doi [\mnras] {10.1093/mnras/203.3.595}, \href
  {https://ui.adsabs.harvard.edu/abs/1983MNRAS.203..595C} {203, 595}

\bibitem[\protect\citeauthoryear{{Chandra}, {Chevalier}, {Chugai}, {Fransson}
  \& {Soderberg}}{{Chandra} et~al.}{2015}]{chandraetal15}
{Chandra} P.,  {Chevalier} R.~A.,  {Chugai} N.,  {Fransson} C.,   {Soderberg}
  A.~M.,  2015, \mn@doi [\apj] {10.1088/0004-637X/810/1/32}, \href
  {https://ui.adsabs.harvard.edu/abs/2015ApJ...810...32C} {810, 32}

\bibitem[\protect\citeauthoryear{{Chen}, {Su}, {Slane}  \& {Wang}}{{Chen}
  et~al.}{2004}]{Chen04}
{Chen} Y.,  {Su} Y.,  {Slane} P.~O.,   {Wang} Q.~D.,  2004, \mn@doi [\apj]
  {10.1086/425152}, \href
  {https://ui.adsabs.harvard.edu/abs/2004ApJ...616..885C} {616, 885}

\bibitem[\protect\citeauthoryear{{Chen}, {Seward}, {Sun}  \& {Li}}{{Chen}
  et~al.}{2008}]{Chen08}
{Chen} Y.,  {Seward} F.~D.,  {Sun} M.,   {Li} J.-t.,  2008, \mn@doi [\apj]
  {10.1086/525240}, \href
  {https://ui.adsabs.harvard.edu/abs/2008ApJ...676.1040C} {676, 1040}

\bibitem[\protect\citeauthoryear{{Chevalier}}{{Chevalier}}{1977}]{rac77}
{Chevalier} R.~A.,  1977, \mn@doi [\araa]
  {10.1146/annurev.aa.15.090177.001135}, \href
  {https://ui.adsabs.harvard.edu/abs/1977ARA&A..15..175C} {15, 175}

\bibitem[\protect\citeauthoryear{{Chevalier}}{{Chevalier}}{1982a}]{rac82a}
{Chevalier} R.~A.,  1982a, \mn@doi [\apj] {10.1086/160126}, \href
  {https://ui.adsabs.harvard.edu/abs/1982ApJ...258..790C} {258, 790}

\bibitem[\protect\citeauthoryear{{Chevalier}}{{Chevalier}}{1982b}]{rac82b}
{Chevalier} R.~A.,  1982b, \mn@doi [\apj] {10.1086/160167}, \href
  {https://ui.adsabs.harvard.edu/abs/1982ApJ...259..302C} {259, 302}

\bibitem[\protect\citeauthoryear{{Chevalier} \& {Dwarkadas}}{{Chevalier} \&
  {Dwarkadas}}{1995}]{cd95}
{Chevalier} R.~A.,  {Dwarkadas} V.~V.,  1995, \mn@doi [\apjl] {10.1086/309714},
  \href {https://ui.adsabs.harvard.edu/abs/1995ApJ...452L..45C} {452, L45}

\bibitem[\protect\citeauthoryear{{Chevalier} \& {Fransson}}{{Chevalier} \&
  {Fransson}}{1994}]{Chev&Fran94}
{Chevalier} R.~A.,  {Fransson} C.,  1994, \mn@doi [\apj] {10.1086/173557},
  \href {https://ui.adsabs.harvard.edu/abs/1994ApJ...420..268C} {420, 268}

\bibitem[\protect\citeauthoryear{{Chiotellis}, {Kosenko}, {Schure}, {Vink}  \&
  {Kaastra}}{{Chiotellis} et~al.}{2013}]{chiotellisetal13}
{Chiotellis} A.,  {Kosenko} D.,  {Schure} K.~M.,  {Vink} J.,   {Kaastra} J.~S.,
   2013, \mn@doi [\mnras] {10.1093/mnras/stt1406}, \href
  {https://ui.adsabs.harvard.edu/abs/2013MNRAS.435.1659C} {435, 1659}

\bibitem[\protect\citeauthoryear{{Cioffi}, {McKee}  \& {Bertschinger}}{{Cioffi}
  et~al.}{1988}]{cmb88}
{Cioffi} D.~F.,  {McKee} C.~F.,   {Bertschinger} E.,  1988, \mn@doi [\apj]
  {10.1086/166834}, \href
  {https://ui.adsabs.harvard.edu/abs/1988ApJ...334..252C} {334, 252}

\bibitem[\protect\citeauthoryear{{Colgate} \& {White}}{{Colgate} \&
  {White}}{1966}]{Colgate&White66}
{Colgate} S.~A.,  {White} R.~H.,  1966, \mn@doi [\apj] {10.1086/148549}, \href
  {https://ui.adsabs.harvard.edu/abs/1966ApJ...143..626C} {143, 626}

\bibitem[\protect\citeauthoryear{{Combi}, {Albacete-Colombo}  \&
  {Mart{\'\i}}}{{Combi} et~al.}{2008}]{Combi08}
{Combi} J.~A.,  {Albacete-Colombo} J.~F.,   {Mart{\'\i}} J.,  2008, \mn@doi
  [\aap] {10.1051/0004-6361:200810378}, \href
  {https://ui.adsabs.harvard.edu/abs/2008A&A...488L..25C} {488, L25}

\bibitem[\protect\citeauthoryear{{Combi} et~al.,}{{Combi}
  et~al.}{2010}]{Combi10}
{Combi} J.~A.,  et~al., 2010, \mn@doi [\aap] {10.1051/0004-6361/200913735},
  \href {https://ui.adsabs.harvard.edu/abs/2010A&A...522A..50C} {522, A50}

\bibitem[\protect\citeauthoryear{{De Horta}, {Collier}, {Filipovi{\'c}},
  {Crawford}, {Uro{\v{s}}evi{\'c}}, {Stootman}  \& {Tothill}}{{De Horta}
  et~al.}{2013}]{DeHorta13}
{De Horta} A.~Y.,  {Collier} J.~D.,  {Filipovi{\'c}} M.~D.,  {Crawford} E.~J.,
  {Uro{\v{s}}evi{\'c}} D.,  {Stootman} F.~H.,   {Tothill} N.~F.~H.,  2013,
  \mn@doi [\mnras] {10.1093/mnras/sts168}, \href
  {https://ui.adsabs.harvard.edu/abs/2013MNRAS.428.1980D} {428, 1980}

\bibitem[\protect\citeauthoryear{{Desai} et~al.,}{{Desai}
  et~al.}{2010}]{Desai10}
{Desai} K.~M.,  et~al., 2010, \mn@doi [\aj] {10.1088/0004-6256/140/2/584},
  \href {https://ui.adsabs.harvard.edu/abs/2010AJ....140..584D} {140, 584}

\bibitem[\protect\citeauthoryear{{Dewey}, {Dwarkadas}, {Haberl}, {Sturm}  \&
  {Canizares}}{{Dewey} et~al.}{2012}]{deweyetal12}
{Dewey} D.,  {Dwarkadas} V.~V.,  {Haberl} F.,  {Sturm} R.,   {Canizares} C.~R.,
   2012, \mn@doi [\apj] {10.1088/0004-637X/752/2/103}, \href
  {https://ui.adsabs.harvard.edu/abs/2012ApJ...752..103D} {752, 103}

\bibitem[\protect\citeauthoryear{{Dickel}}{{Dickel}}{2020}]{dickel20}
{Dickel} J.~R.,  2020, \mn@doi [\aj] {10.3847/1538-3881/abaccf}, \href
  {https://ui.adsabs.harvard.edu/abs/2020AJ....160..157D} {160, 157}

\bibitem[\protect\citeauthoryear{{Dubner}, {Moffett}, {Goss}  \&
  {Winkler}}{{Dubner} et~al.}{1993}]{Dubner93}
{Dubner} G.~M.,  {Moffett} D.~A.,  {Goss} W.~M.,   {Winkler} P.~F.,  1993,
  \mn@doi [\aj] {10.1086/116603}, \href
  {https://ui.adsabs.harvard.edu/abs/1993AJ....105.2251D} {105, 2251}

\bibitem[\protect\citeauthoryear{{Dubner}, {Giacani}, {Goss}, {Moffett}  \&
  {Holdaway}}{{Dubner} et~al.}{1996}]{Dubner96}
{Dubner} G.~M.,  {Giacani} E.~B.,  {Goss} W.~M.,  {Moffett} D.~A.,   {Holdaway}
  M.,  1996, \mn@doi [\aj] {10.1086/117875}, \href
  {https://ui.adsabs.harvard.edu/abs/1996AJ....111.1304D} {111, 1304}

\bibitem[\protect\citeauthoryear{{Dubner}, {Giacani}, {Goss}, {Green}  \&
  {Nyman}}{{Dubner} et~al.}{2002}]{Dubner02}
{Dubner} G.~M.,  {Giacani} E.~B.,  {Goss} W.~M.,  {Green} A.~J.,   {Nyman}
  L.~{\r{A}}.,  2002, \mn@doi [\aap] {10.1051/0004-6361:20020365}, \href
  {https://ui.adsabs.harvard.edu/abs/2002A&A...387.1047D} {387, 1047}

\bibitem[\protect\citeauthoryear{{Dubner}, {Loiseau}, {Rodr{\'\i}guez-Pascual},
  {Smith}, {Giacani}  \& {Castelletti}}{{Dubner} et~al.}{2013}]{Dubner13}
{Dubner} G.,  {Loiseau} N.,  {Rodr{\'\i}guez-Pascual} P.,  {Smith} M.~J.~S.,
  {Giacani} E.,   {Castelletti} G.,  2013, \mn@doi [\aap]
  {10.1051/0004-6361/201321401}, \href
  {https://ui.adsabs.harvard.edu/abs/2013A&A...555A...9D} {555, A9}

\bibitem[\protect\citeauthoryear{{Dwarkadas}}{{Dwarkadas}}{2005}]{Dwarkadas05}
{Dwarkadas} V.~V.,  2005, \mn@doi [\apj] {10.1086/432109}, \href
  {https://ui.adsabs.harvard.edu/abs/2005ApJ...630..892D} {630, 892}

\bibitem[\protect\citeauthoryear{{Dwarkadas}}{{Dwarkadas}}{2011}]{dwarkadas11}
{Dwarkadas} V.~V.,  2011, \memsai, \href
  {https://ui.adsabs.harvard.edu/abs/2011MmSAI..82..781D} {82, 781}

\bibitem[\protect\citeauthoryear{{Dwarkadas}}{{Dwarkadas}}{2014}]{vvd14}
{Dwarkadas} V.~V.,  2014, \mn@doi [\mnras] {10.1093/mnras/stu347}, \href
  {https://ui.adsabs.harvard.edu/abs/2014MNRAS.440.1917D} {440, 1917}

\bibitem[\protect\citeauthoryear{{Dwarkadas} \& {Chevalier}}{{Dwarkadas} \&
  {Chevalier}}{1998}]{Dwarkadas&Chev98}
{Dwarkadas} V.~V.,  {Chevalier} R.~A.,  1998, \mn@doi [\apj] {10.1086/305478},
  \href {https://ui.adsabs.harvard.edu/abs/1998ApJ...497..807D} {497, 807}

\bibitem[\protect\citeauthoryear{{Dwarkadas} \& {Gruszko}}{{Dwarkadas} \&
  {Gruszko}}{2012}]{dg12}
{Dwarkadas} V.~V.,  {Gruszko} J.,  2012, \mn@doi [\mnras]
  {10.1111/j.1365-2966.2011.19808.x}, \href
  {https://ui.adsabs.harvard.edu/abs/2012MNRAS.419.1515D} {419, 1515}

\bibitem[\protect\citeauthoryear{{Dwarkadas}, {Bauer}, {Bietenholz}  \&
  {Bartel}}{{Dwarkadas} et~al.}{2014}]{dbbb14}
{Dwarkadas} V.,  {Bauer} F.,  {Bietenholz} M.,   {Bartel} N.,  2014, in {Ness}
  J.-U.,  ed., The X-ray Universe 2014. p.~248

\bibitem[\protect\citeauthoryear{{Dwarkadas}, {Romero-Ca{\~n}izales}, {Reddy}
  \& {Bauer}}{{Dwarkadas} et~al.}{2016}]{drrb16}
{Dwarkadas} V.~V.,  {Romero-Ca{\~n}izales} C.,  {Reddy} R.,   {Bauer} F.~E.,
  2016, \mn@doi [\mnras] {10.1093/mnras/stw1717}, \href
  {https://ui.adsabs.harvard.edu/abs/2016MNRAS.462.1101D} {462, 1101}

\bibitem[\protect\citeauthoryear{{Ferrand} \& {Safi-Harb}}{{Ferrand} \&
  {Safi-Harb}}{2012}]{FerrandSafi-Harb12}
{Ferrand} G.,  {Safi-Harb} S.,  2012, \mn@doi [Advances in Space Research]
  {10.1016/j.asr.2012.02.004}, \href
  {https://ui.adsabs.harvard.edu/abs/2012AdSpR..49.1313F} {49, 1313}

\bibitem[\protect\citeauthoryear{{Finkelstein} et~al.,}{{Finkelstein}
  et~al.}{2006}]{Finkelstein06}
{Finkelstein} S.~L.,  et~al., 2006, \mn@doi [\apj] {10.1086/500570}, \href
  {https://ui.adsabs.harvard.edu/abs/2006ApJ...641..919F} {641, 919}

\bibitem[\protect\citeauthoryear{{Fox} et~al.,}{{Fox} et~al.}{2014}]{foxetal14}
{Fox} O.~D.,  et~al., 2014, \mn@doi [\apj] {10.1088/0004-637X/790/1/17}, \href
  {https://ui.adsabs.harvard.edu/abs/2014ApJ...790...17F} {790, 17}

\bibitem[\protect\citeauthoryear{{Frail}, {Goss}, {Reynoso}, {Giacani}, {Green}
   \& {Otrupcek}}{{Frail} et~al.}{1996}]{Frail96}
{Frail} D.~A.,  {Goss} W.~M.,  {Reynoso} E.~M.,  {Giacani} E.~B.,  {Green}
  A.~J.,   {Otrupcek} R.,  1996, \mn@doi [\aj] {10.1086/117904}, \href
  {https://ui.adsabs.harvard.edu/abs/1996AJ....111.1651F} {111, 1651}

\bibitem[\protect\citeauthoryear{{Frank}, {Zhekov}, {Park}, {McCray}, {Dwek}
  \& {Burrows}}{{Frank} et~al.}{2016}]{Frank16}
{Frank} K.~A.,  {Zhekov} S.~A.,  {Park} S.,  {McCray} R.,  {Dwek} E.,
  {Burrows} D.~N.,  2016, \mn@doi [\apj] {10.3847/0004-637X/829/1/40}, \href
  {https://ui.adsabs.harvard.edu/abs/2016ApJ...829...40F} {829, 40}

\bibitem[\protect\citeauthoryear{{Frank}, {Dwarkadas}, {Panfichi}, {Crum}  \&
  {Burrows}}{{Frank} et~al.}{2019}]{franketal19}
{Frank} K.~A.,  {Dwarkadas} V.,  {Panfichi} A.,  {Crum} R.~M.,   {Burrows}
  D.~N.,  2019, \mn@doi [\apj] {10.3847/1538-4357/ab0e81}, \href
  {https://ui.adsabs.harvard.edu/abs/2019ApJ...875...14F} {875, 14}

\bibitem[\protect\citeauthoryear{{Fransson}, {Lundqvist}  \&
  {Chevalier}}{{Fransson} et~al.}{1996}]{flc96}
{Fransson} C.,  {Lundqvist} P.,   {Chevalier} R.~A.,  1996, \mn@doi [\apj]
  {10.1086/177119}, \href
  {https://ui.adsabs.harvard.edu/abs/1996ApJ...461..993F} {461, 993}

\bibitem[\protect\citeauthoryear{{Fransson} et~al.,}{{Fransson}
  et~al.}{2014}]{franssonetal14}
{Fransson} C.,  et~al., 2014, \mn@doi [\apj] {10.1088/0004-637X/797/2/118},
  \href {https://ui.adsabs.harvard.edu/abs/2014ApJ...797..118F} {797, 118}

\bibitem[\protect\citeauthoryear{{Fransson} et~al.,}{{Fransson}
  et~al.}{2015}]{franssonetal15}
{Fransson} C.,  et~al., 2015, \mn@doi [\apjl] {10.1088/2041-8205/806/1/L19},
  \href {https://ui.adsabs.harvard.edu/abs/2015ApJ...806L..19F} {806, L19}

\bibitem[\protect\citeauthoryear{{Gaensler}, {Pivovaroff}  \&
  {Garmire}}{{Gaensler} et~al.}{2001}]{Gaensler01}
{Gaensler} B.~M.,  {Pivovaroff} M.~J.,   {Garmire} G.~P.,  2001, \mn@doi
  [\apjl] {10.1086/322982}, \href
  {https://ui.adsabs.harvard.edu/abs/2001ApJ...556L.107G} {556, L107}

\bibitem[\protect\citeauthoryear{{Gaensler} et~al.,}{{Gaensler}
  et~al.}{2008}]{Gaensler08}
{Gaensler} B.~M.,  et~al., 2008, \mn@doi [\apjl] {10.1086/589650}, \href
  {https://ui.adsabs.harvard.edu/abs/2008ApJ...680L..37G} {680, L37}

\bibitem[\protect\citeauthoryear{{Gaetz}, {Butt}, {Edgar}, {Eriksen},
  {Plucinsky}, {Schlegel}  \& {Smith}}{{Gaetz} et~al.}{2000}]{Gaetz00}
{Gaetz} T.~J.,  {Butt} Y.~M.,  {Edgar} R.~J.,  {Eriksen} K.~A.,  {Plucinsky}
  P.~P.,  {Schlegel} E.~M.,   {Smith} R.~K.,  2000, \mn@doi [\apjl]
  {10.1086/312640}, \href
  {https://ui.adsabs.harvard.edu/abs/2000ApJ...534L..47G} {534, L47}

\bibitem[\protect\citeauthoryear{{Garc{\'\i}a}, {Combi}, {Albacete-Colombo},
  {Romero}, {Bocchino}  \& {L{\'o}pez-Santiago}}{{Garc{\'\i}a}
  et~al.}{2012}]{Garcia12}
{Garc{\'\i}a} F.,  {Combi} J.~A.,  {Albacete-Colombo} J.~F.,  {Romero} G.~E.,
  {Bocchino} F.,   {L{\'o}pez-Santiago} J.,  2012, \mn@doi [\aap]
  {10.1051/0004-6361/201218959}, \href
  {https://ui.adsabs.harvard.edu/abs/2012A&A...546A..91G} {546, A91}

\bibitem[\protect\citeauthoryear{{Ghavamian} \& {Williams}}{{Ghavamian} \&
  {Williams}}{2016}]{GhavamianWilliams16}
{Ghavamian} P.,  {Williams} B.~J.,  2016, \mn@doi [\apj]
  {10.3847/0004-637X/831/2/188}, \href
  {https://ui.adsabs.harvard.edu/abs/2016ApJ...831..188G} {831, 188}

\bibitem[\protect\citeauthoryear{{Ghavamian}, {Rakowski}, {Hughes}  \&
  {Williams}}{{Ghavamian} et~al.}{2003}]{ghavamianetal03}
{Ghavamian} P.,  {Rakowski} C.~E.,  {Hughes} J.~P.,   {Williams} T.~B.,  2003,
  \mn@doi [\apj] {10.1086/375161}, \href
  {https://ui.adsabs.harvard.edu/abs/2003ApJ...590..833G} {590, 833}

\bibitem[\protect\citeauthoryear{{Ghavamian}, {Laming}  \&
  {Rakowski}}{{Ghavamian} et~al.}{2007}]{glr07}
{Ghavamian} P.,  {Laming} J.~M.,   {Rakowski} C.~E.,  2007, \mn@doi [\apjl]
  {10.1086/510740}, \href
  {https://ui.adsabs.harvard.edu/abs/2007ApJ...654L..69G} {654, L69}

\bibitem[\protect\citeauthoryear{{Giacani}, {Smith}, {Dubner}, {Loiseau},
  {Castelletti}  \& {Paron}}{{Giacani} et~al.}{2009}]{Giacani09}
{Giacani} E.,  {Smith} M.~J.~S.,  {Dubner} G.,  {Loiseau} N.,  {Castelletti}
  G.,   {Paron} S.,  2009, \mn@doi [\aap] {10.1051/0004-6361/200912253}, \href
  {https://ui.adsabs.harvard.edu/abs/2009A&A...507..841G} {507, 841}

\bibitem[\protect\citeauthoryear{{Giacani}, {Smith}, {Dubner}  \&
  {Loiseau}}{{Giacani} et~al.}{2011}]{Giacani11}
{Giacani} E.,  {Smith} M.~J.~S.,  {Dubner} G.,   {Loiseau} N.,  2011, \mn@doi
  [\aap] {10.1051/0004-6361/201116768}, \href
  {https://ui.adsabs.harvard.edu/abs/2011A&A...531A.138G} {531, A138}

\bibitem[\protect\citeauthoryear{{Giordano} et~al.,}{{Giordano}
  et~al.}{2012}]{Giordano12}
{Giordano} F.,  et~al., 2012, \mn@doi [\apjl] {10.1088/2041-8205/744/1/L2},
  \href {https://ui.adsabs.harvard.edu/abs/2012ApJ...744L...2G} {744, L2}

\bibitem[\protect\citeauthoryear{{Gotthelf} \& {Vasisht}}{{Gotthelf} \&
  {Vasisht}}{1997}]{GotthelfVasisht97}
{Gotthelf} E.~V.,  {Vasisht} G.,  1997, \mn@doi [\apjl] {10.1086/310846}, \href
  {https://ui.adsabs.harvard.edu/abs/1997ApJ...486L.133G} {486, L133}

\bibitem[\protect\citeauthoryear{{Green}}{{Green}}{1984}]{Green84}
{Green} D.~A.,  1984, \mn@doi [\mnras] {10.1093/mnras/209.3.449}, \href
  {https://ui.adsabs.harvard.edu/abs/1984MNRAS.209..449G} {209, 449}

\bibitem[\protect\citeauthoryear{{Green}, {Gull}, {Tan}  \& {Simon}}{{Green}
  et~al.}{1988}]{Green88}
{Green} D.~A.,  {Gull} S.~F.,  {Tan} S.~M.,   {Simon} A.~J.~B.,  1988, \mn@doi
  [\mnras] {10.1093/mnras/231.3.735}, \href
  {https://ui.adsabs.harvard.edu/abs/1988MNRAS.231..735G} {231, 735}

\bibitem[\protect\citeauthoryear{{Greiner}, {Egger}  \& {Aschenbach}}{{Greiner}
  et~al.}{1994}]{Greiner94}
{Greiner} J.,  {Egger} R.,   {Aschenbach} B.,  1994, \aap, \href
  {https://ui.adsabs.harvard.edu/abs/1994A&A...286L..35G} {286, L35}

\bibitem[\protect\citeauthoryear{{Gull}}{{Gull}}{1973}]{gull73}
{Gull} S.~F.,  1973, \mn@doi [\mnras] {10.1093/mnras/161.1.47}, \href
  {https://ui.adsabs.harvard.edu/abs/1973MNRAS.161...47G} {161, 47}

\bibitem[\protect\citeauthoryear{{Harrus}, {Slane}, {Smith}  \&
  {Hughes}}{{Harrus} et~al.}{2001}]{Harrus01}
{Harrus} I.~M.,  {Slane} P.~O.,  {Smith} R.~K.,   {Hughes} J.~P.,  2001,
  \mn@doi [\apj] {10.1086/320577}, \href
  {https://ui.adsabs.harvard.edu/abs/2001ApJ...552..614H} {552, 614}

\bibitem[\protect\citeauthoryear{{Hendrick}, {Reynolds}  \&
  {Borkowski}}{{Hendrick} et~al.}{2005}]{Hendrick05}
{Hendrick} S.~P.,  {Reynolds} S.~P.,   {Borkowski} K.~J.,  2005, \mn@doi
  [\apjl] {10.1086/429862}, \href
  {https://ui.adsabs.harvard.edu/abs/2005ApJ...622L.117H} {622, L117}

\bibitem[\protect\citeauthoryear{{Huang}, {Wu}, {Hui}, {Seo}, {Trepl}  \&
  {Kong}}{{Huang} et~al.}{2014}]{Huang14}
{Huang} R.~H.~H.,  {Wu} J.~H.~K.,  {Hui} C.~Y.,  {Seo} K.~A.,  {Trepl} L.,
  {Kong} A.~K.~H.,  2014, \mn@doi [\apj] {10.1088/0004-637X/785/2/118}, \href
  {https://ui.adsabs.harvard.edu/abs/2014ApJ...785..118H} {785, 118}

\bibitem[\protect\citeauthoryear{{Hughes}, {Ghavamian}, {Rakowski}  \&
  {Slane}}{{Hughes} et~al.}{2003}]{hughesetal03}
{Hughes} J.~P.,  {Ghavamian} P.,  {Rakowski} C.~E.,   {Slane} P.~O.,  2003,
  \mn@doi [\apjl] {10.1086/367760}, \href
  {https://ui.adsabs.harvard.edu/abs/2003ApJ...582L..95H} {582, L95}

\bibitem[\protect\citeauthoryear{{Hui} \& {Becker}}{{Hui} \&
  {Becker}}{2009}]{HuiBecker09}
{Hui} C.~Y.,  {Becker} W.,  2009, \mn@doi [\aap] {10.1051/0004-6361:200810789},
  \href {https://ui.adsabs.harvard.edu/abs/2009A&A...494.1005H} {494, 1005}

\bibitem[\protect\citeauthoryear{{Hwang} \& {Gotthelf}}{{Hwang} \&
  {Gotthelf}}{1997}]{hg97}
{Hwang} U.,  {Gotthelf} E.~V.,  1997, \mn@doi [\apj] {10.1086/303546}, \href
  {https://ui.adsabs.harvard.edu/abs/1997ApJ...475..665H} {475, 665}

\bibitem[\protect\citeauthoryear{{Hwang} \& {Laming}}{{Hwang} \&
  {Laming}}{2012}]{hl12}
{Hwang} U.,  {Laming} J.~M.,  2012, \mn@doi [\apj]
  {10.1088/0004-637X/746/2/130}, \href
  {https://ui.adsabs.harvard.edu/abs/2012ApJ...746..130H} {746, 130}

\bibitem[\protect\citeauthoryear{{Hwang}, {Flanagan}  \& {Petre}}{{Hwang}
  et~al.}{2005}]{Hwang05}
{Hwang} U.,  {Flanagan} K.~A.,   {Petre} R.,  2005, \mn@doi [\apj]
  {10.1086/497298}, \href
  {https://ui.adsabs.harvard.edu/abs/2005ApJ...635..355H} {635, 355}

\bibitem[\protect\citeauthoryear{{Janka}}{{Janka}}{2012}]{Janka12}
{Janka} H.-T.,  2012, \mn@doi [Annual Review of Nuclear and Particle Science]
  {10.1146/annurev-nucl-102711-094901}, \href
  {https://ui.adsabs.harvard.edu/abs/2012ARNPS..62..407J} {62, 407}

\bibitem[\protect\citeauthoryear{{Jones} et~al.,}{{Jones}
  et~al.}{1998}]{Jones98}
{Jones} T.~W.,  et~al., 1998, \mn@doi [\pasp] {10.1086/316122}, \href
  {https://ui.adsabs.harvard.edu/abs/1998PASP..110..125J} {110, 125}

\bibitem[\protect\citeauthoryear{{Kim}, {Rieke}, {Krause}, {Misselt},
  {Indebetouw}  \& {Johnson}}{{Kim} et~al.}{2008}]{Kim08}
{Kim} Y.,  {Rieke} G.~H.,  {Krause} O.,  {Misselt} K.,  {Indebetouw} R.,
  {Johnson} K.~E.,  2008, \mn@doi [\apj] {10.1086/533426}, \href
  {https://ui.adsabs.harvard.edu/abs/2008ApJ...678..287K} {678, 287}

\bibitem[\protect\citeauthoryear{{Kinugasa}, {Torii}, {Tsunemi}, {Yamauchi},
  {Koyama}  \& {Dotani}}{{Kinugasa} et~al.}{1998}]{Kinugasa98}
{Kinugasa} K.,  {Torii} K.,  {Tsunemi} H.,  {Yamauchi} S.,  {Koyama} K.,
  {Dotani} T.,  1998, \mn@doi [\pasj] {10.1093/pasj/50.2.249}, \href
  {https://ui.adsabs.harvard.edu/abs/1998PASJ...50..249K} {50, 249}

\bibitem[\protect\citeauthoryear{{Koo}, {Moon}, {Lee}, {Lee}  \&
  {Matthews}}{{Koo} et~al.}{2007}]{Koo07}
{Koo} B.-C.,  {Moon} D.-S.,  {Lee} H.-G.,  {Lee} J.-J.,   {Matthews} K.,  2007,
  \mn@doi [\apj] {10.1086/510550}, \href
  {https://ui.adsabs.harvard.edu/abs/2007ApJ...657..308K} {657, 308}

\bibitem[\protect\citeauthoryear{{Krause}, {Birkmann}, {Usuda}, {Hattori},
  {Goto}, {Rieke}  \& {Misselt}}{{Krause} et~al.}{2008}]{krauseetal08}
{Krause} O.,  {Birkmann} S.~M.,  {Usuda} T.,  {Hattori} T.,  {Goto} M.,
  {Rieke} G.~H.,   {Misselt} K.~A.,  2008, \mn@doi [Science]
  {10.1126/science.1155788}, \href
  {https://ui.adsabs.harvard.edu/abs/2008Sci...320.1195K} {320, 1195}

\bibitem[\protect\citeauthoryear{{Laming} \& {Hwang}}{{Laming} \&
  {Hwang}}{2003}]{LamingHwang03}
{Laming} J.~M.,  {Hwang} U.,  2003, \mn@doi [\apj] {10.1086/378268}, \href
  {https://ui.adsabs.harvard.edu/abs/2003ApJ...597..347L} {597, 347}

\bibitem[\protect\citeauthoryear{{Lazendic}, {Slane}, {Hughes}, {Chen}  \&
  {Dame}}{{Lazendic} et~al.}{2005}]{Lazendic05}
{Lazendic} J.~S.,  {Slane} P.~O.,  {Hughes} J.~P.,  {Chen} Y.,   {Dame} T.~M.,
  2005, \mn@doi [\apj] {10.1086/426114}, \href
  {https://ui.adsabs.harvard.edu/abs/2005ApJ...618..733L} {618, 733}

\bibitem[\protect\citeauthoryear{{Leahy} \& {Ranasinghe}}{{Leahy} \&
  {Ranasinghe}}{2016}]{LeahyRanasinghe16}
{Leahy} D.~A.,  {Ranasinghe} S.,  2016, \mn@doi [\apj]
  {10.3847/0004-637X/817/1/74}, \href
  {https://ui.adsabs.harvard.edu/abs/2016ApJ...817...74L} {817, 74}

\bibitem[\protect\citeauthoryear{{Lee}, {Moon}, {Koo}, {Lee}  \&
  {Matthews}}{{Lee} et~al.}{2009}]{Lee09}
{Lee} H.-G.,  {Moon} D.-S.,  {Koo} B.-C.,  {Lee} J.-J.,   {Matthews} K.,  2009,
  \mn@doi [\apj] {10.1088/0004-637X/691/2/1042}, \href
  {https://ui.adsabs.harvard.edu/abs/2009ApJ...691.1042L} {691, 1042}

\bibitem[\protect\citeauthoryear{{Lee}, {Park}, {Hughes}  \& {Slane}}{{Lee}
  et~al.}{2014}]{leeetal14}
{Lee} J.-J.,  {Park} S.,  {Hughes} J.~P.,   {Slane} P.~O.,  2014, \mn@doi
  [\apj] {10.1088/0004-637X/789/1/7}, \href
  {https://ui.adsabs.harvard.edu/abs/2014ApJ...789....7L} {789, 7}

\bibitem[\protect\citeauthoryear{{Leitherer}, {Robert}  \&
  {Drissen}}{{Leitherer} et~al.}{1992}]{Leitherer92}
{Leitherer} C.,  {Robert} C.,   {Drissen} L.,  1992, \mn@doi [\apj]
  {10.1086/172089}, \href
  {https://ui.adsabs.harvard.edu/abs/1992ApJ...401..596L} {401, 596}

\bibitem[\protect\citeauthoryear{{Lopez}, {Ramirez-Ruiz}, {Castro}  \&
  {Pearson}}{{Lopez} et~al.}{2013}]{lopezetal13}
{Lopez} L.~A.,  {Ramirez-Ruiz} E.,  {Castro} D.,   {Pearson} S.,  2013, \mn@doi
  [\apj] {10.1088/0004-637X/764/1/50}, \href
  {https://ui.adsabs.harvard.edu/abs/2013ApJ...764...50L} {764, 50}

\bibitem[\protect\citeauthoryear{{Lovchinsky}, {Slane}, {Gaensler}, {Hughes},
  {Ng}, {Lazendic}, {Gelfand}  \& {Brogan}}{{Lovchinsky}
  et~al.}{2011}]{Lovchinsky11}
{Lovchinsky} I.,  {Slane} P.,  {Gaensler} B.~M.,  {Hughes} J.~P.,  {Ng} C.~Y.,
  {Lazendic} J.~S.,  {Gelfand} J.~D.,   {Brogan} C.~L.,  2011, \mn@doi [\apj]
  {10.1088/0004-637X/731/1/70}, \href
  {https://ui.adsabs.harvard.edu/abs/2011ApJ...731...70L} {731, 70}

\bibitem[\protect\citeauthoryear{{Luo} \& {McCray}}{{Luo} \&
  {McCray}}{1991}]{lm91}
{Luo} D.,  {McCray} R.,  1991, \mn@doi [\apj] {10.1086/170539}, \href
  {https://ui.adsabs.harvard.edu/abs/1991ApJ...379..659L} {379, 659}

\bibitem[\protect\citeauthoryear{{Maggi} et~al.,}{{Maggi}
  et~al.}{2016}]{Maggi16}
{Maggi} P.,  et~al., 2016, \mn@doi [\aap] {10.1051/0004-6361/201526932}, \href
  {https://ui.adsabs.harvard.edu/abs/2016A&A...585A.162M} {585, A162}

\bibitem[\protect\citeauthoryear{{Maggi} et~al.,}{{Maggi}
  et~al.}{2019}]{Maggi19}
{Maggi} P.,  et~al., 2019, \mn@doi [\aap] {10.1051/0004-6361/201936583}, \href
  {https://ui.adsabs.harvard.edu/abs/2019A&A...631A.127M} {631, A127}

\bibitem[\protect\citeauthoryear{{Mart{\'\i}nez-Rodr{\'\i}guez}
  et~al.,}{{Mart{\'\i}nez-Rodr{\'\i}guez} et~al.}{2020}]{martinezetal20}
{Mart{\'\i}nez-Rodr{\'\i}guez} H.,  et~al., 2020, arXiv e-prints, \href
  {https://ui.adsabs.harvard.edu/abs/2020arXiv200608681M} {p. arXiv:2006.08681}

\bibitem[\protect\citeauthoryear{{Matteucci} \& {Francois}}{{Matteucci} \&
  {Francois}}{1989}]{Matteucci&Francois89}
{Matteucci} F.,  {Francois} P.,  1989, \mn@doi [\mnras]
  {10.1093/mnras/239.3.885}, \href
  {https://ui.adsabs.harvard.edu/abs/1989MNRAS.239..885M} {239, 885}

\bibitem[\protect\citeauthoryear{{Maund}, {Smartt}, {Kudritzki},
  {Podsiadlowski}  \& {Gilmore}}{{Maund} et~al.}{2004}]{maundetal04}
{Maund} J.~R.,  {Smartt} S.~J.,  {Kudritzki} R.~P.,  {Podsiadlowski} P.,
  {Gilmore} G.~F.,  2004, \mn@doi [\nat] {10.1038/nature02161}, \href
  {https://ui.adsabs.harvard.edu/abs/2004Natur.427..129M} {427, 129}

\bibitem[\protect\citeauthoryear{{Maxted} et~al.,}{{Maxted}
  et~al.}{2013}]{Maxted13}
{Maxted} N.~I.,  et~al., 2013, \mn@doi [\mnras] {10.1093/mnras/stt1159}, \href
  {https://ui.adsabs.harvard.edu/abs/2013MNRAS.434.2188M} {434, 2188}

\bibitem[\protect\citeauthoryear{{McCray} \& {Fransson}}{{McCray} \&
  {Fransson}}{2016}]{mf16}
{McCray} R.,  {Fransson} C.,  2016, \mn@doi [\araa]
  {10.1146/annurev-astro-082615-105405}, \href
  {https://ui.adsabs.harvard.edu/abs/2016ARA&A..54...19M} {54, 19}

\bibitem[\protect\citeauthoryear{{McEntaffer}, {Grieves}, {DeRoo}  \&
  {Brantseg}}{{McEntaffer} et~al.}{2013}]{McEntaffer13}
{McEntaffer} R.~L.,  {Grieves} N.,  {DeRoo} C.,   {Brantseg} T.,  2013, \mn@doi
  [\apj] {10.1088/0004-637X/774/2/120}, \href
  {https://ui.adsabs.harvard.edu/abs/2013ApJ...774..120M} {774, 120}

\bibitem[\protect\citeauthoryear{{McKee} \& {Ostriker}}{{McKee} \&
  {Ostriker}}{1977}]{mo77}
{McKee} C.~F.,  {Ostriker} J.~P.,  1977, \mn@doi [\apj] {10.1086/155667}, \href
  {https://ui.adsabs.harvard.edu/abs/1977ApJ...218..148M} {218, 148}

\bibitem[\protect\citeauthoryear{{Mereghetti}, {Sidoli}  \&
  {Israel}}{{Mereghetti} et~al.}{1998}]{Mereghetti98}
{Mereghetti} S.,  {Sidoli} L.,   {Israel} G.~L.,  1998, \aap, \href
  {https://ui.adsabs.harvard.edu/abs/1998A&A...331L..77M} {331, L77}

\bibitem[\protect\citeauthoryear{{Miceli}, {Decourchelle}, {Ballet},
  {Bocchino}, {Hughes}, {Hwang}  \& {Petre}}{{Miceli} et~al.}{2008}]{Miceli08}
{Miceli} M.,  {Decourchelle} A.,  {Ballet} J.,  {Bocchino} F.,  {Hughes} J.,
  {Hwang} U.,   {Petre} R.,  2008, \mn@doi [Advances in Space Research]
  {10.1016/j.asr.2007.01.030}, \href
  {https://ui.adsabs.harvard.edu/abs/2008AdSpR..41..390M} {41, 390}

\bibitem[\protect\citeauthoryear{{Miceli}, {Sciortino}, {Troja}  \&
  {Orlando}}{{Miceli} et~al.}{2015}]{micelietal15}
{Miceli} M.,  {Sciortino} S.,  {Troja} E.,   {Orlando} S.,  2015, \mn@doi
  [\apj] {10.1088/0004-637X/805/2/120}, \href
  {https://ui.adsabs.harvard.edu/abs/2015ApJ...805..120M} {805, 120}

\bibitem[\protect\citeauthoryear{{Morlino} \& {Caprioli}}{{Morlino} \&
  {Caprioli}}{2012}]{MorlinoCaprioli12}
{Morlino} G.,  {Caprioli} D.,  2012, \mn@doi [\aap]
  {10.1051/0004-6361/201117855}, \href
  {https://ui.adsabs.harvard.edu/abs/2012A&A...538A..81M} {538, A81}

\bibitem[\protect\citeauthoryear{{Nomoto}}{{Nomoto}}{1982}]{Nomoto82}
{Nomoto} K.,  1982, \mn@doi [\apj] {10.1086/159682}, \href
  {https://ui.adsabs.harvard.edu/abs/1982ApJ...253..798N} {253, 798}

\bibitem[\protect\citeauthoryear{{Olbert}, {Keohane}, {Arnaud}, {Dyer},
  {Reynolds}  \& {Safi-Harb}}{{Olbert} et~al.}{2003}]{Olbert03}
{Olbert} C.~M.,  {Keohane} J.~W.,  {Arnaud} K.~A.,  {Dyer} K.~K.,  {Reynolds}
  S.~P.,   {Safi-Harb} S.,  2003, \mn@doi [\apjl] {10.1086/377348}, \href
  {https://ui.adsabs.harvard.edu/abs/2003ApJ...592L..45O} {592, L45}

\bibitem[\protect\citeauthoryear{{Ostriker} \& {McKee}}{{Ostriker} \&
  {McKee}}{1988}]{om88}
{Ostriker} J.~P.,  {McKee} C.~F.,  1988, \mn@doi [Reviews of Modern Physics]
  {10.1103/RevModPhys.60.1}, \href
  {https://ui.adsabs.harvard.edu/abs/1988RvMP...60....1O} {60, 1}

\bibitem[\protect\citeauthoryear{{Ou}, {Chu}, {Maggi}, {Li}, {Chang}  \&
  {Gruendl}}{{Ou} et~al.}{2018}]{Ou18}
{Ou} P.-S.,  {Chu} Y.-H.,  {Maggi} P.,  {Li} C.-J.,  {Chang} U.~P.,   {Gruendl}
  R.~A.,  2018, \mn@doi [\apj] {10.3847/1538-4357/aad04b}, \href
  {https://ui.adsabs.harvard.edu/abs/2018ApJ...863..137O} {863, 137}

\bibitem[\protect\citeauthoryear{{Pannuti}, {Kargaltsev}, {Napier}  \&
  {Brehm}}{{Pannuti} et~al.}{2014}]{Pannuti14}
{Pannuti} T.~G.,  {Kargaltsev} O.,  {Napier} J.~P.,   {Brehm} D.,  2014,
  \mn@doi [\apj] {10.1088/0004-637X/782/2/102}, \href
  {https://ui.adsabs.harvard.edu/abs/2014ApJ...782..102P} {782, 102}

\bibitem[\protect\citeauthoryear{{Park}, {Roming}, {Hughes}, {Slane},
  {Burrows}, {Garmire}  \& {Nousek}}{{Park} et~al.}{2002}]{Park02}
{Park} S.,  {Roming} P. W.~A.,  {Hughes} J.~P.,  {Slane} P.~O.,  {Burrows}
  D.~N.,  {Garmire} G.~P.,   {Nousek} J.~A.,  2002, \mn@doi [\apjl]
  {10.1086/338861}, \href
  {https://ui.adsabs.harvard.edu/abs/2002ApJ...564L..39P} {564, L39}

\bibitem[\protect\citeauthoryear{{Park}, {Hughes}, {Burrows}, {Slane}, {Nousek}
   \& {Garmire}}{{Park} et~al.}{2003}]{Park03}
{Park} S.,  {Hughes} J.~P.,  {Burrows} D.~N.,  {Slane} P.~O.,  {Nousek} J.~A.,
   {Garmire} G.~P.,  2003, \mn@doi [\apjl] {10.1086/380599}, \href
  {https://ui.adsabs.harvard.edu/abs/2003ApJ...598L..95P} {598, L95}

\bibitem[\protect\citeauthoryear{{Park}, {Kargaltsev}, {Pavlov}, {Mori},
  {Slane}, {Hughes}, {Burrows}  \& {Garmire}}{{Park} et~al.}{2009}]{Park09}
{Park} S.,  {Kargaltsev} O.,  {Pavlov} G.~G.,  {Mori} K.,  {Slane} P.~O.,
  {Hughes} J.~P.,  {Burrows} D.~N.,   {Garmire} G.~P.,  2009, \mn@doi [\apj]
  {10.1088/0004-637X/695/1/431}, \href
  {https://ui.adsabs.harvard.edu/abs/2009ApJ...695..431P} {695, 431}

\bibitem[\protect\citeauthoryear{{Patnaude}, {Badenes}, {Park}  \&
  {Laming}}{{Patnaude} et~al.}{2012}]{patnaudeteal12}
{Patnaude} D.~J.,  {Badenes} C.,  {Park} S.,   {Laming} J.~M.,  2012, \mn@doi
  [\apj] {10.1088/0004-637X/756/1/6}, \href
  {https://ui.adsabs.harvard.edu/abs/2012ApJ...756....6P} {756, 6}

\bibitem[\protect\citeauthoryear{{Prinz} \& {Becker}}{{Prinz} \&
  {Becker}}{2012}]{PrinzBecker12}
{Prinz} T.,  {Becker} W.,  2012, \mn@doi [\aap] {10.1051/0004-6361/201219086},
  \href {https://ui.adsabs.harvard.edu/abs/2012A&A...544A...7P} {544, A7}

\bibitem[\protect\citeauthoryear{{Rakowski}, {Badenes}, {Gaensler}, {Gelfand},
  {Hughes}  \& {Slane}}{{Rakowski} et~al.}{2006}]{Rakowski06}
{Rakowski} C.~E.,  {Badenes} C.,  {Gaensler} B.~M.,  {Gelfand} J.~D.,  {Hughes}
  J.~P.,   {Slane} P.~O.,  2006, \mn@doi [\apj] {10.1086/505018}, \href
  {https://ui.adsabs.harvard.edu/abs/2006ApJ...646..982R} {646, 982}

\bibitem[\protect\citeauthoryear{{Ramakrishnan} \& {Dwarkadas}}{{Ramakrishnan}
  \& {Dwarkadas}}{2020}]{rd20}
{Ramakrishnan} V.,  {Dwarkadas} V.~V.,  2020, \mn@doi [\apj]
  {10.3847/1538-4357/abb087}, \href
  {https://ui.adsabs.harvard.edu/abs/2020ApJ...901..119R} {901, 119}

\bibitem[\protect\citeauthoryear{{Raymond}, {Cox}  \& {Smith}}{{Raymond}
  et~al.}{1976}]{Raymond76}
{Raymond} J.~C.,  {Cox} D.~P.,   {Smith} B.~W.,  1976, \mn@doi [\apj]
  {10.1086/154170}, \href
  {https://ui.adsabs.harvard.edu/abs/1976ApJ...204..290R} {204, 290}

\bibitem[\protect\citeauthoryear{{Reed}, {Hester}, {Fabian}  \&
  {Winkler}}{{Reed} et~al.}{1995}]{Reed95}
{Reed} J.~E.,  {Hester} J.~J.,  {Fabian} A.~C.,   {Winkler} P.~F.,  1995,
  \mn@doi [\apj] {10.1086/175308}, \href
  {https://ui.adsabs.harvard.edu/abs/1995ApJ...440..706R} {440, 706}

\bibitem[\protect\citeauthoryear{{Reynolds}, {Borkowski}, {Hwang}, {Harrus},
  {Petre}  \& {Dubner}}{{Reynolds} et~al.}{2006}]{Reynolds06}
{Reynolds} S.~P.,  {Borkowski} K.~J.,  {Hwang} U.,  {Harrus} I.,  {Petre} R.,
  {Dubner} G.,  2006, \mn@doi [\apjl] {10.1086/510066}, \href
  {https://ui.adsabs.harvard.edu/abs/2006ApJ...652L..45R} {652, L45}

\bibitem[\protect\citeauthoryear{{Reynolds}, {Borkowski}, {Green}, {Hwang},
  {Harrus}  \& {Petre}}{{Reynolds} et~al.}{2008}]{Reynolds08}
{Reynolds} S.~P.,  {Borkowski} K.~J.,  {Green} D.~A.,  {Hwang} U.,  {Harrus}
  I.,   {Petre} R.,  2008, \mn@doi [\apjl] {10.1086/589570}, \href
  {https://ui.adsabs.harvard.edu/abs/2008ApJ...680L..41R} {680, L41}

\bibitem[\protect\citeauthoryear{{Reynolds} et~al.,}{{Reynolds}
  et~al.}{2013}]{Reynolds13}
{Reynolds} M.~T.,  et~al., 2013, \mn@doi [\apj] {10.1088/0004-637X/766/2/112},
  \href {https://ui.adsabs.harvard.edu/abs/2013ApJ...766..112R} {766, 112}

\bibitem[\protect\citeauthoryear{{Reynoso} \& {Goss}}{{Reynoso} \&
  {Goss}}{1999}]{Reynoso99}
{Reynoso} E.~M.,  {Goss} W.~M.,  1999, \mn@doi [\aj] {10.1086/300990}, \href
  {https://ui.adsabs.harvard.edu/abs/1999AJ....118..926R} {118, 926}

\bibitem[\protect\citeauthoryear{{Ross} \& {Dwarkadas}}{{Ross} \&
  {Dwarkadas}}{2017}]{RossDwarkadas17}
{Ross} M.,  {Dwarkadas} V.~V.,  2017, \mn@doi [\aj] {10.3847/1538-3881/aa6d50},
  \href {https://ui.adsabs.harvard.edu/abs/2017AJ....153..246R} {153, 246}

\bibitem[\protect\citeauthoryear{{Safi-Harb}, {Dubner}, {Petre}, {Holt}  \&
  {Durouchoux}}{{Safi-Harb} et~al.}{2005}]{Safi-Harb05}
{Safi-Harb} S.,  {Dubner} G.,  {Petre} R.,  {Holt} S.~S.,   {Durouchoux} P.,
  2005, \mn@doi [\apj] {10.1086/425960}, \href
  {https://ui.adsabs.harvard.edu/abs/2005ApJ...618..321S} {618, 321}

\bibitem[\protect\citeauthoryear{{Sasaki}, {M{\"a}kel{\"a}}, {Klochkov},
  {Santangelo}  \& {Suleimanov}}{{Sasaki} et~al.}{2018}]{Sasaki18}
{Sasaki} M.,  {M{\"a}kel{\"a}} M.~M.,  {Klochkov} D.,  {Santangelo} A.,
  {Suleimanov} V.,  2018, \mn@doi [\mnras] {10.1093/mnras/sty1596}, \href
  {https://ui.adsabs.harvard.edu/abs/2018MNRAS.479.3033S} {479, 3033}

\bibitem[\protect\citeauthoryear{{Sato}, {Koyama}, {Takahashi}, {Odaka}  \&
  {Nakashima}}{{Sato} et~al.}{2014}]{Sato14}
{Sato} T.,  {Koyama} K.,  {Takahashi} T.,  {Odaka} H.,   {Nakashima} S.,  2014,
  \mn@doi [\pasj] {10.1093/pasj/psu120}, \href
  {https://ui.adsabs.harvard.edu/abs/2014PASJ...66..124S} {66, 124}

\bibitem[\protect\citeauthoryear{{Sawada}, {Tachibana}, {Uchida}, {Ito},
  {Matsumura}, {Bamba}, {Tsuru}  \& {Tanaka}}{{Sawada} et~al.}{2019}]{Sawada19}
{Sawada} M.,  {Tachibana} K.,  {Uchida} H.,  {Ito} Y.,  {Matsumura} H.,
  {Bamba} A.,  {Tsuru} T.~G.,   {Tanaka} T.,  2019, \mn@doi [\pasj]
  {10.1093/pasj/psz036}, \href
  {https://ui.adsabs.harvard.edu/abs/2019PASJ...71...61S} {71, 61}

\bibitem[\protect\citeauthoryear{{Schenck}, {Park}, {Burrows}, {Hughes}, {Lee}
  \& {Mori}}{{Schenck} et~al.}{2014}]{Schenck14}
{Schenck} A.,  {Park} S.,  {Burrows} D.~N.,  {Hughes} J.~P.,  {Lee} J.-J.,
  {Mori} K.,  2014, \mn@doi [\apj] {10.1088/0004-637X/791/1/50}, \href
  {https://ui.adsabs.harvard.edu/abs/2014ApJ...791...50S} {791, 50}

\bibitem[\protect\citeauthoryear{{Schlegel}}{{Schlegel}}{1995}]{Schlegel95}
{Schlegel} E.~M.,  1995, \mn@doi [Reports on Progress in Physics]
  {10.1088/0034-4885/58/11/001}, \href
  {https://ui.adsabs.harvard.edu/abs/1995RPPh...58.1375S} {58, 1375}

\bibitem[\protect\citeauthoryear{{Sedov}}{{Sedov}}{1959}]{Sedov59}
{Sedov} L.~I.,  1959, {Similarity and Dimensional Methods in Mechanics}.
{New York: Academic Press}

\bibitem[\protect\citeauthoryear{{Seward} \& {Charles}}{{Seward} \&
  {Charles}}{2010}]{SewardCharles10}
{Seward} F.~D.,  {Charles} P.~A.,  2010, {Exploring the X-ray Universe}.
{Cambridge University Press}

\bibitem[\protect\citeauthoryear{{Seward}, {Kearns}  \& {Rhode}}{{Seward}
  et~al.}{1996}]{Seward96}
{Seward} F.~D.,  {Kearns} K.~E.,   {Rhode} K.~L.,  1996, \mn@doi [\apj]
  {10.1086/178015}, \href
  {https://ui.adsabs.harvard.edu/abs/1996ApJ...471..887S} {471, 887}

\bibitem[\protect\citeauthoryear{{Sidoli}, {Bocchino}, {Mereghetti}  \&
  {Bandiera}}{{Sidoli} et~al.}{2004}]{Sidoli04}
{Sidoli} L.,  {Bocchino} F.,  {Mereghetti} S.,   {Bandiera} R.,  2004, \memsai,
  \href {https://ui.adsabs.harvard.edu/abs/2004MmSAI..75..507S} {75, 507}

\bibitem[\protect\citeauthoryear{{Siegel}, {Dwarkadas}, {Frank}  \&
  {Burrows}}{{Siegel} et~al.}{2020}]{siegeletal20b}
{Siegel} J.,  {Dwarkadas} V.~V.,  {Frank} K.~A.,   {Burrows} D.~N.,  2020,
  arXiv e-prints, \href {https://ui.adsabs.harvard.edu/abs/2020arXiv201004765S}
  {p. arXiv:2010.04765}

\bibitem[\protect\citeauthoryear{{Siegel}, {Dwarkadas}, {Frank}  \&
  {Burrows}}{{Siegel} et~al.}{2021}]{siegeletal21}
{Siegel} J.,  {Dwarkadas} V.~V.,  {Frank} K.~A.,   {Burrows} D.~N.,  2021,
  arXiv e-prints, \href {https://ui.adsabs.harvard.edu/abs/2021arXiv210901157S}
  {p. arXiv:2109.01157}

\bibitem[\protect\citeauthoryear{{Slane}, {Smith}, {Hughes}  \&
  {Petre}}{{Slane} et~al.}{2002}]{Slane02}
{Slane} P.,  {Smith} R.~K.,  {Hughes} J.~P.,   {Petre} R.,  2002, \mn@doi
  [\apj] {10.1086/324155}, \href
  {https://ui.adsabs.harvard.edu/abs/2002ApJ...564..284S} {564, 284}

\bibitem[\protect\citeauthoryear{{Su} \& {Chen}}{{Su} \&
  {Chen}}{2005}]{SuChen05}
{Su} Y.,  {Chen} Y.,  2005, \mn@doi [\cjaa] {10.1088/1009-9271/5/4/008}, \href
  {https://ui.adsabs.harvard.edu/abs/2005ChJAA...5..412S} {5, 412}

\bibitem[\protect\citeauthoryear{{Sukhbold}, {Ertl}, {Woosley}, {Brown}  \&
  {Janka}}{{Sukhbold} et~al.}{2016}]{sukhboldetal16}
{Sukhbold} T.,  {Ertl} T.,  {Woosley} S.~E.,  {Brown} J.~M.,   {Janka} H.~T.,
  2016, \mn@doi [\apj] {10.3847/0004-637X/821/1/38}, \href
  {https://ui.adsabs.harvard.edu/abs/2016ApJ...821...38S} {821, 38}

\bibitem[\protect\citeauthoryear{{Sun} \& {Chen}}{{Sun} \&
  {Chen}}{2019}]{SunChen19}
{Sun} L.,  {Chen} Y.,  2019, \mn@doi [\apj] {10.3847/1538-4357/aafb73}, \href
  {https://ui.adsabs.harvard.edu/abs/2019ApJ...872...45S} {872, 45}

\bibitem[\protect\citeauthoryear{{Sun}, {Seward}, {Smith}  \& {Slane}}{{Sun}
  et~al.}{2004}]{Sun04}
{Sun} M.,  {Seward} F.~D.,  {Smith} R.~K.,   {Slane} P.~O.,  2004, \mn@doi
  [\apj] {10.1086/382666}, \href
  {https://ui.adsabs.harvard.edu/abs/2004ApJ...605..742S} {605, 742}

\bibitem[\protect\citeauthoryear{{Taylor}}{{Taylor}}{1950}]{Taylor50}
{Taylor} G.,  1950, \mn@doi [Proceedings of the Royal Society of London Series
  A] {10.1098/rspa.1950.0049}, \href
  {https://ui.adsabs.harvard.edu/abs/1950RSPSA.201..159T} {201, 159}

\bibitem[\protect\citeauthoryear{{Temim}, {Slane}, {Kolb}, {Blondin}, {Hughes}
  \& {Bucciantini}}{{Temim} et~al.}{2015}]{Temim15}
{Temim} T.,  {Slane} P.,  {Kolb} C.,  {Blondin} J.,  {Hughes} J.~P.,
  {Bucciantini} N.,  2015, \mn@doi [\apj] {10.1088/0004-637X/808/1/100}, \href
  {https://ui.adsabs.harvard.edu/abs/2015ApJ...808..100T} {808, 100}

\bibitem[\protect\citeauthoryear{{Truelove} \& {McKee}}{{Truelove} \&
  {McKee}}{1999}]{Truelove&Mckee99}
{Truelove} J.~K.,  {McKee} C.~F.,  1999, \mn@doi [\apjs] {10.1086/313176},
  \href {https://ui.adsabs.harvard.edu/abs/1999ApJS..120..299T} {120, 299}

\bibitem[\protect\citeauthoryear{{Truran}, {Arnett}  \& {Cameron}}{{Truran}
  et~al.}{1967}]{Truran67}
{Truran} J.~W.,  {Arnett} W.~D.,   {Cameron} A.~G.~W.,  1967, \mn@doi [Canadian
  Journal of Physics] {10.1139/p67-184}, \href
  {https://ui.adsabs.harvard.edu/abs/1967CaJPh..45.2315T} {45, 2315}

\bibitem[\protect\citeauthoryear{{Uchida}, {Yamaguchi}  \& {Koyama}}{{Uchida}
  et~al.}{2013}]{Uchida13}
{Uchida} H.,  {Yamaguchi} H.,   {Koyama} K.,  2013, \mn@doi [\apj]
  {10.1088/0004-637X/771/1/56}, \href
  {https://ui.adsabs.harvard.edu/abs/2013ApJ...771...56U} {771, 56}

\bibitem[\protect\citeauthoryear{{Vink}}{{Vink}}{2004}]{Vink04}
{Vink} J.,  2004, \mn@doi [\apj] {10.1086/381930}, \href
  {https://ui.adsabs.harvard.edu/abs/2004ApJ...604..693V} {604, 693}

\bibitem[\protect\citeauthoryear{{Vink}}{{Vink}}{2012}]{Vink12}
{Vink} J.,  2012, \mn@doi [\aapr] {10.1007/s00159-011-0049-1}, \href
  {https://ui.adsabs.harvard.edu/abs/2012A&ARv..20...49V} {20, 49}

\bibitem[\protect\citeauthoryear{{Vink}}{{Vink}}{2020}]{Vink20}
{Vink} J.,  2020, {Physics and Evolution of Supernova Remnants}.
{Springer Nature Switzerland AG}

\bibitem[\protect\citeauthoryear{{Vink} \& {Sander}}{{Vink} \&
  {Sander}}{2021}]{vs21}
{Vink} J.~S.,  {Sander} A. A.~C.,  2021, \mn@doi [\mnras]
  {10.1093/mnras/stab902}, \href
  {https://ui.adsabs.harvard.edu/abs/2021MNRAS.504.2051V} {504, 2051}

\bibitem[\protect\citeauthoryear{{Vink}, {de Koter}  \& {Lamers}}{{Vink}
  et~al.}{2001}]{vinketal01}
{Vink} J.~S.,  {de Koter} A.,   {Lamers} H.~J.~G.~L.~M.,  2001, \mn@doi [\aap]
  {10.1051/0004-6361:20010127}, \href
  {https://ui.adsabs.harvard.edu/abs/2001A&A...369..574V} {369, 574}

\bibitem[\protect\citeauthoryear{{Warren} et~al.,}{{Warren}
  et~al.}{2005}]{Warren05}
{Warren} J.~S.,  et~al., 2005, \mn@doi [\apj] {10.1086/496941}, \href
  {https://ui.adsabs.harvard.edu/abs/2005ApJ...634..376W} {634, 376}

\bibitem[\protect\citeauthoryear{{Weaver}, {McCray}, {Castor}, {Shapiro}  \&
  {Moore}}{{Weaver} et~al.}{1977}]{weaveretal77}
{Weaver} R.,  {McCray} R.,  {Castor} J.,  {Shapiro} P.,   {Moore} R.,  1977,
  \mn@doi [\apj] {10.1086/155692}, \href
  {https://ui.adsabs.harvard.edu/abs/1977ApJ...218..377W} {218, 377}

\bibitem[\protect\citeauthoryear{{Webbink}}{{Webbink}}{1984}]{Webbink84}
{Webbink} R.~F.,  1984, \mn@doi [\apj] {10.1086/161701}, \href
  {https://ui.adsabs.harvard.edu/abs/1984ApJ...277..355W} {277, 355}

\bibitem[\protect\citeauthoryear{{Wilhelm}, {Telezhinsky}, {Dwarkadas}  \&
  {Pohl}}{{Wilhelm} et~al.}{2020}]{wtdp20}
{Wilhelm} A.,  {Telezhinsky} I.,  {Dwarkadas} V.~V.,   {Pohl} M.,  2020,
  \mn@doi [\aap] {10.1051/0004-6361/201936079}, \href
  {https://ui.adsabs.harvard.edu/abs/2020A&A...639A.124W} {639, A124}

\bibitem[\protect\citeauthoryear{{Williams}, {Hewitt}, {Petre}  \&
  {Temim}}{{Williams} et~al.}{2018}]{Williams18}
{Williams} B.~J.,  {Hewitt} J.~W.,  {Petre} R.,   {Temim} T.,  2018, \mn@doi
  [\apj] {10.3847/1538-4357/aaadb6}, \href
  {https://ui.adsabs.harvard.edu/abs/2018ApJ...855..118W} {855, 118}

\bibitem[\protect\citeauthoryear{{Williams}, {Katsuda}, {Cumbee}, {Petre},
  {Raymond}  \& {Uchida}}{{Williams} et~al.}{2020}]{williamsetal20}
{Williams} B.~J.,  {Katsuda} S.,  {Cumbee} R.,  {Petre} R.,  {Raymond} J.~C.,
  {Uchida} H.,  2020, \mn@doi [\apjl] {10.3847/2041-8213/aba7c1}, \href
  {https://ui.adsabs.harvard.edu/abs/2020ApJ...898L..51W} {898, L51}

\bibitem[\protect\citeauthoryear{{Wilms}, {Allen}  \& {McCray}}{{Wilms}
  et~al.}{2000}]{Wilms00}
{Wilms} J.,  {Allen} A.,   {McCray} R.,  2000, \mn@doi [\apj] {10.1086/317016},
  \href {https://ui.adsabs.harvard.edu/abs/2000ApJ...542..914W} {542, 914}

\bibitem[\protect\citeauthoryear{{Winkler}, {Twelker}, {Reith}  \&
  {Long}}{{Winkler} et~al.}{2009}]{Winkler09}
{Winkler} P.~F.,  {Twelker} K.,  {Reith} C.~N.,   {Long} K.~S.,  2009, \mn@doi
  [\apj] {10.1088/0004-637X/692/2/1489}, \href
  {https://ui.adsabs.harvard.edu/abs/2009ApJ...692.1489W} {692, 1489}

\bibitem[\protect\citeauthoryear{{Woltjer}}{{Woltjer}}{1972}]{Woltjer72}
{Woltjer} L.,  1972, \mn@doi [\araa] {10.1146/annurev.aa.10.090172.001021},
  \href {https://ui.adsabs.harvard.edu/abs/1972ARA&A..10..129W} {10, 129}

\bibitem[\protect\citeauthoryear{{Woosley} \& {Kasen}}{{Woosley} \&
  {Kasen}}{2011}]{WoosleyKasen11}
{Woosley} S.~E.,  {Kasen} D.,  2011, \mn@doi [\apj]
  {10.1088/0004-637X/734/1/38}, \href
  {https://ui.adsabs.harvard.edu/abs/2011ApJ...734...38W} {734, 38}

\bibitem[\protect\citeauthoryear{{Yamaguchi}, {Ozawa}  \&
  {Ohnishi}}{{Yamaguchi} et~al.}{2012}]{Yamaguchi12}
{Yamaguchi} H.,  {Ozawa} M.,   {Ohnishi} T.,  2012, \mn@doi [Advances in Space
  Research] {10.1016/j.asr.2011.11.002}, \href
  {https://ui.adsabs.harvard.edu/abs/2012AdSpR..49..451Y} {49, 451}

\bibitem[\protect\citeauthoryear{{Yamaguchi} et~al.,}{{Yamaguchi}
  et~al.}{2014a}]{yamaguchietal14}
{Yamaguchi} H.,  et~al., 2014a, \mn@doi [\apj] {10.1088/0004-637X/780/2/136},
  \href {https://ui.adsabs.harvard.edu/abs/2014ApJ...780..136Y} {780, 136}

\bibitem[\protect\citeauthoryear{{Yamaguchi} et~al.,}{{Yamaguchi}
  et~al.}{2014b}]{Yamaguchi_Badenes14}
{Yamaguchi} H.,  et~al., 2014b, \mn@doi [\apjl] {10.1088/2041-8205/785/2/L27},
  \href {https://ui.adsabs.harvard.edu/abs/2014ApJ...785L..27Y} {785, L27}

\bibitem[\protect\citeauthoryear{{Yamaguchi}, {Hughes}, {Badenes}, {Bravo},
  {Seitenzahl}, {Mart{\'\i}nez-Rodr{\'\i}guez}, {Park}  \& {Petre}}{{Yamaguchi}
  et~al.}{2017}]{yamaguchietal17}
{Yamaguchi} H.,  {Hughes} J.~P.,  {Badenes} C.,  {Bravo} E.,  {Seitenzahl}
  I.~R.,  {Mart{\'\i}nez-Rodr{\'\i}guez} H.,  {Park} S.,   {Petre} R.,  2017,
  \mn@doi [\apj] {10.3847/1538-4357/834/2/124}, \href
  {https://ui.adsabs.harvard.edu/abs/2017ApJ...834..124Y} {834, 124}

\bibitem[\protect\citeauthoryear{{Yamaguchi}, {Acero}, {Li}  \&
  {Chu}}{{Yamaguchi} et~al.}{2021}]{yamaguchietal21}
{Yamaguchi} H.,  {Acero} F.,  {Li} C.-J.,   {Chu} Y.-H.,  2021, \mn@doi [\apjl]
  {10.3847/2041-8213/abee8a}, \href
  {https://ui.adsabs.harvard.edu/abs/2021ApJ...910L..24Y} {910, L24}

\bibitem[\protect\citeauthoryear{{Yamauchi}, {Minami}, {Ota}  \&
  {Koyama}}{{Yamauchi} et~al.}{2014}]{Yamauchi_Minami14}
{Yamauchi} S.,  {Minami} S.,  {Ota} N.,   {Koyama} K.,  2014, \mn@doi [\pasj]
  {10.1093/pasj/pst004}, \href
  {https://ui.adsabs.harvard.edu/abs/2014PASJ...66....2Y} {66, 2}

\bibitem[\protect\citeauthoryear{{Yasumi}, {Nobukawa}, {Nakashima}, {Uchida},
  {Sugawara}, {Tsuru}, {Tanaka}  \& {Koyama}}{{Yasumi} et~al.}{2014}]{Yasumi14}
{Yasumi} M.,  {Nobukawa} M.,  {Nakashima} S.,  {Uchida} H.,  {Sugawara} R.,
  {Tsuru} T.~G.,  {Tanaka} T.,   {Koyama} K.,  2014, \mn@doi [\pasj]
  {10.1093/pasj/psu043}, \href
  {https://ui.adsabs.harvard.edu/abs/2014PASJ...66...68Y} {66, 68}

\bibitem[\protect\citeauthoryear{{Yew} et~al.,}{{Yew} et~al.}{2021}]{Yew21}
{Yew} M.,  et~al., 2021, \mn@doi [\mnras] {10.1093/mnras/staa3382}, \href
  {https://ui.adsabs.harvard.edu/abs/2021MNRAS.500.2336Y} {500, 2336}

\bibitem[\protect\citeauthoryear{{Yokogawa}, {Imanishi}, {Koyama}, {Nishiuchi}
  \& {Mizuno}}{{Yokogawa} et~al.}{2002}]{Yokogawa02}
{Yokogawa} J.,  {Imanishi} K.,  {Koyama} K.,  {Nishiuchi} M.,   {Mizuno} N.,
  2002, \mn@doi [\pasj] {10.1093/pasj/54.1.53}, \href
  {https://ui.adsabs.harvard.edu/abs/2002PASJ...54...53Y} {54, 53}

\bibitem[\protect\citeauthoryear{{Zhang}, {Chen}, {Su}, {Zhou}, {Pannuti}  \&
  {Zhou}}{{Zhang} et~al.}{2015}]{Zhang15}
{Zhang} G.-Y.,  {Chen} Y.,  {Su} Y.,  {Zhou} X.,  {Pannuti} T.~G.,   {Zhou} P.,
   2015, \mn@doi [\apj] {10.1088/0004-637X/799/1/103}, \href
  {https://ui.adsabs.harvard.edu/abs/2015ApJ...799..103Z} {799, 103}

\bibitem[\protect\citeauthoryear{{Zhou} \& {Vink}}{{Zhou} \&
  {Vink}}{2018}]{ZhouVink18}
{Zhou} P.,  {Vink} J.,  2018, \mn@doi [\aap] {10.1051/0004-6361/201731583},
  \href {https://ui.adsabs.harvard.edu/abs/2018A&A...615A.150Z} {615, A150}

\bibitem[\protect\citeauthoryear{{Zhou}, {Chen}, {Safi-Harb}, {Zhou}, {Sun},
  {Zhang}  \& {Zhang}}{{Zhou} et~al.}{2016}]{Zhou16}
{Zhou} P.,  {Chen} Y.,  {Safi-Harb} S.,  {Zhou} X.,  {Sun} M.,  {Zhang} Z.-Y.,
   {Zhang} G.-Y.,  2016, \mn@doi [\apj] {10.3847/0004-637X/831/2/192}, \href
  {https://ui.adsabs.harvard.edu/abs/2016ApJ...831..192Z} {831, 192}

\bibitem[\protect\citeauthoryear{{Zhu}, {Tian}  \& {Zuo}}{{Zhu}
  et~al.}{2014}]{Zhu14}
{Zhu} H.,  {Tian} W.~W.,   {Zuo} P.,  2014, \mn@doi [\apj]
  {10.1088/0004-637X/793/2/95}, \href
  {https://ui.adsabs.harvard.edu/abs/2014ApJ...793...95Z} {793, 95}

\bibitem[\protect\citeauthoryear{{van Kerkwijk}, {Chang}  \& {Justham}}{{van
  Kerkwijk} et~al.}{2010}]{Kerkwijk10}
{van Kerkwijk} M.~H.,  {Chang} P.,   {Justham} S.,  2010, \mn@doi [\apjl]
  {10.1088/2041-8205/722/2/L157}, \href
  {https://ui.adsabs.harvard.edu/abs/2010ApJ...722L.157V} {722, L157}

\bibitem[\protect\citeauthoryear{{van der Heyden}, {Bleeker}, {Kaastra}  \&
  {Vink}}{{van der Heyden} et~al.}{2003}]{vanderheydenetal03}
{van der Heyden} K.~J.,  {Bleeker} J.~A.~M.,  {Kaastra} J.~S.,   {Vink} J.,
  2003, \mn@doi [\aap] {10.1051/0004-6361:20030658}, \href
  {https://ui.adsabs.harvard.edu/abs/2003A&A...406..141V} {406, 141}

\bibitem[\protect\citeauthoryear{{van der Heyden}, {Bleeker}  \&
  {Kaastra}}{{van der Heyden} et~al.}{2004}]{Heyden04}
{van der Heyden} K.~J.,  {Bleeker} J.~A.~M.,   {Kaastra} J.~S.,  2004, \mn@doi
  [\aap] {10.1051/0004-6361:20034156}, \href
  {https://ui.adsabs.harvard.edu/abs/2004A&A...421.1031V} {421, 1031}

\makeatother
\end{thebibliography}
\end{multicols}
\end{document}